\documentclass[a4paper,abstract,toc=bib]{scrreprt}
\usepackage{scrhack}
\title{Analyzing and Evaluating the Behavior of Git Diff and Merge}
\author{Niels Glodny}
\date{\today}
\usepackage[backend=biber,style=numeric,sorting=none]{biblatex}

\usepackage{amsmath}
\usepackage{microtype}
\usepackage{hyperref}
\usepackage{cleveref}
\usepackage{subcaption}
\usepackage{algorithm2e}
\Crefname{algocf}{Alg.}{Algs.}
\Crefname{algocf}{Algorithm}{Algorithms}
\usepackage{amstext}
\usepackage{booktabs}
\usepackage{setspace}

\usepackage{listings}
\usepackage{adjustbox}
\usepackage[svgnames]{xcolor}
\definecolor{diffstart}{named}{Grey}
\definecolor{diffincl}{named}{Green}
\definecolor{diffrem}{named}{OrangeRed}
\definecolor{conflict}{named}{Blue}

  \lstdefinelanguage{diff}{
    basicstyle=\ttfamily\small,
      morecomment=[s][\color{diffstart}]{@@}{@@},
      morecomment=[f][\color{diffincl}]{+},
      morecomment=[f][\color{diffrem}]{-},
      morecomment=[s][\color{conflict}]{<<<}{>>>},
      morecomment=[l][\color{conflict}]{>>>},
  }

\DeclareBibliographyAlias{letter}{online}

\usepackage{hyperref}

\usepackage{tikz}
\usetikzlibrary{arrows.meta, math, calc, positioning, decorations.pathreplacing}
\usepackage{ifthen}
\usepackage{stmaryrd}

\usepackage{xstring}
\newcommand{\commithash}[1]{\texttt{{#1}}}
\newcommand{\sourcecite}[2]{\ifx&#2&[\texttt{#1}]\else[\texttt{#1}, l.\,#2]\fi}
\newcommand{\commitcite}[1]{\href{https://github.com/git/git/commit/#1}{[commit\nobreak\,\,\commithash{#1}]}}
\newcommand{\commitsourcecite}[3]{%
  \href{
    https://github.com/git/git/blob/#1/#2
  }{[commit \commithash{#1}, file \texttt{#2}, l.\,#3]}%
}

\usepackage{enumitem}

\begin{document}
\pagenumbering{roman}
\begin{titlepage}
  
\begin{center}
    \uppercase{Ludwig-Maximilians-Universität München}
\end{center}
\begin{center}
    \uppercase{Institut für Informatik}
\end{center}

\vspace*{10mm}
\begin{center}
    \includegraphics[height=40mm]{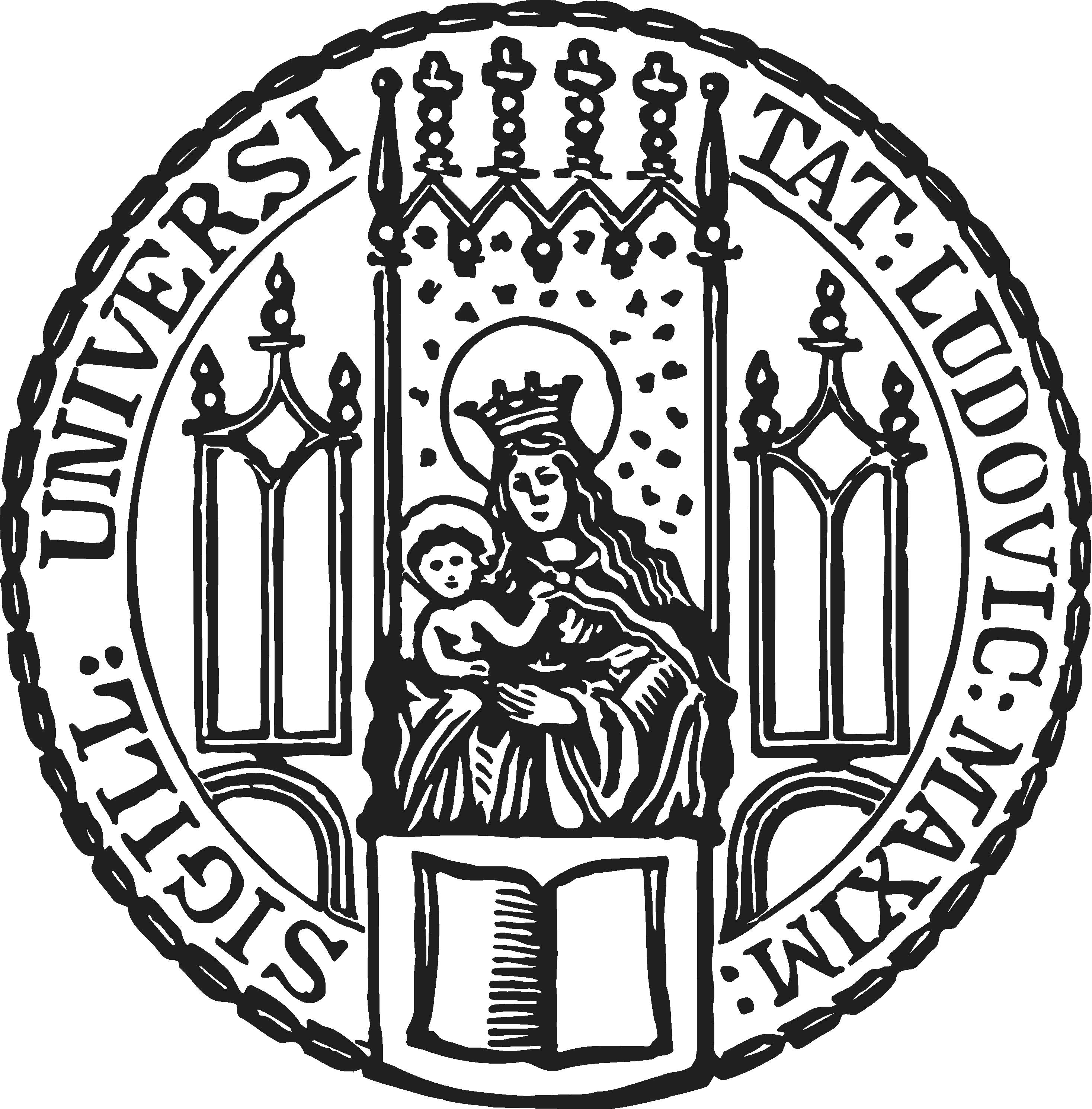}
\end{center}

\vspace*{10mm}
\date{\vspace{-3ex}}
\begin{center}
    {\huge Analyzing and Evaluating the Behavior of Git\\ 
    \vspace{0.15cm}
    Diff and Merge}
    
    \vspace{0.5cm}

    {\large Niels Glodny}
\end{center}
\thispagestyle{empty}
\vspace{1cm}
\begin{center}
    \begin{Large}
        Bachelor's Thesis\\
    \end{Large}
    \vspace{0.2cm}
    \begin{large}
        in Computer Science
    \end{large}
\end{center}
\vspace{3cm}
\begin{center}
        \begin{tabular}{ll}
            Supervisors: & Dr. Martin Kleppmann  (University of Cambridge) \\
                         & Prof. Dr. Gidon Ernst (Ludwig-Maximilians-Universität München)\\
                         &\\
                         &\\
            Submission Date:& July 4, 2025
        \end{tabular}
\end{center}
\vspace{1,5cm}

\end{titlepage}
\newpage
% \thispagestyle{empty}
% \vspace*{\fill}
% \noindent \textbf{Statement of Originality}
% \\
% \\\noindent
% I hereby confirm that I have written the accompanying thesis by myself, without
% contributions from any sources other than those cited in the text and acknowledgments. 
% I have only used generative AI to assist in writing code for the empirical evaluations.
% \\
% \vspace{1cm}

% \hspace*{\fill}\includegraphics{graphics/signature.pdf}\\
% \noindent Munich, July 4, 2025 \hfill Niels Glodny

% \newpage
\begin{abstract}
Despite being widely used, the algorithms that enable collaboration with Git are not well understood.
The diff and merge algorithms are particularly interesting, as they could be applied in other contexts.
In this thesis, I document the main functionalities of Git: how diffs are computed, how they are used to run merges, and how merges enable more complex operations.
In the process, I show multiple unexpected behaviors in Git, including the following:
The histogram diff algorithm has pathological cases where a single-line change can cause the entire rest of the file to be marked as changed.
The default merge strategy (\texttt{ort}) can result in merges requiring exponential time in the number of commits in the history.
Merges and rebases are not commutative, and even when merges do not result in a conflict, the result is not specified but depends on the diff algorithm used.
And finally, sometimes when two sides of a merge add different lines at the same position, the result is not a conflict, but a merge containing both changes after each other, in arbitrary order.
\end{abstract}

\tableofcontents
\chapter{Introduction}
\pagenumbering{arabic}

In a modern software engineering workflow, the use of version control systems (VCS) is ubiquitous. 
They are used to store the source code and its history as well as to synchronize when working together with other developers. 
One of the earliest such systems, the Revision Control System (RCS) \cite{tichyRCSSystemVersion1985}, started as a set of scripts on top of UNIX utilities.
It was not collaborative and only designed to track files individually.
From there on, a history of systems has emerged, always trying to improve upon the tools that were popular at the time.
The Concurrent Versions System (CVS) improved upon RCS by making it possible to collaborate over the network with a client-server architecture \cite{rupareliaHistoryVersionControl2010a}.
Today, Git \cite{Git} is the most widely used VCS by a large margin, with over 90\% of developers reporting to use it \cite{StackOverflow2020}.
Besides Git, other approaches to version control have been developed as well.
Many of these systems have in common that they are \textit{distributed}, meaning that each user has a local copy of the entire history, on which they can perform VCS operations without depending on a server.
This creates the need for \textit{merging}, combining changes that were created independently of each other.

Git has developed organically from the rich heritage of version control systems and was never designed with a theoretical model behind it.
Instead, it started as a tool for storing source code and added many of its version control features around it.
Many of the text based collaboration algorithms used in Git have been developed over time and are based on heuristics that proved useful in practice.

For example, the patience and histogram algorithms have been developed for Git, with very little description on their motivation, properties, or operation. 
Multiple heuristics have been added on top of otherwise known algorithms, improving their behavior in certain practical situations, but altering the guarantees that the algorithm would provide without the heuristics.
Surprisingly little research has been done on these algorithms and heuristics, despite their widespread use in Git by millions of developers.

\paragraph{Related Work}
Projects like Darcs \cite{DarcsnetTheory} or Pijul \cite{pijulauthorsPijul} have tried to create a version control system where operations like branching and merging are formally well-understood.
In the case of Darcs, this is done through a theory of patches, which has been analyzed in detail \cite{mimramCategoricalTheoryPatches2013,angiuliHomotopicalPatchTheory2014}.
These systems are based on storing changes between the versions and manipulating these changes, whereas Git stores a snapshot of file contents at each commit.

Outside the area of version control software, other systems for distributed and collaborative editing of text have been developed, like Operational Transformation \cite{sunOperationalTransformationRealtime1998} and Conflict-Free Replicated Data Types (CRDTs) \cite{shapiroConflictFreeReplicatedData2011}.
Both of these techniques are well-studied and can give guarantees for the result of a merging operation.

While the three-way merge algorithm has been described before \cite{khannaFormalInvestigationDiff32007}, it is only a small part of the algorithms that make collaborative text editing with Git possible.
Other important parts--in particular the algorithms for comparing two files--have neither been analyzed nor even described outside the source code.
Before the development of Git started, there had been a body of research focusing on comparing files \cite{wagnerStringtoStringCorrectionProblem1974,hirschbergLinearSpaceAlgorithm1975}, with Myers algorithm \cite{myersONDDifferenceAlgorithm1986} being widely adopted in practice.
Further related work around particular algorithms is discussed within the sections of this thesis that explain those algorithms.

\paragraph{Contributions}
The goal of this work is to give a comprehensive overview of the algorithms used in Git for comparing files and merging changes and to analyze their properties.
These are not only interesting to gain a better understanding of the most widely deployed version control system, but might also have the potential to be incorporated into other text-based collaboration systems, including collaborative document editors such as Google Docs and Overleaf, which currently lack support for version control concepts such as branching.
I have structured this thesis into three parts.
The first part will give an overview of the architecture of Git and how its components work together to enable tools like merging, rebasing, cherry-picking, and reverting.
The second part focuses on the algorithms for file comparison in Git and the third part focuses on the merging of files.
Both the high-level merge strategies and the underlying three-way merge algorithm will be discussed.
While some parts of the system are already understood, I have filled the gaps with my own research based on both analyzing the source code and running experiments.
Besides the overview of Git and its algorithms, I have made the following main contributions:
\begin{itemize}
    \item I provide the first detailed description and motivation for the \texttt{histogram} diff algorithm in Git, demonstrating that, contrary to previous claims, it is not simply an extension of the \texttt{patience} algorithm.
    \item I conduct an empirical analysis of the size of diffs generated by the diff algorithms in Git on real-world commits, which shows that the patience and histogram algorithms give overall similar results that are, on average, slightly larger-than-optimal.
    \item I develop a technique for constructing adversarial diffs where the \texttt{histogram} algorithm performs highly suboptimally, marking the entire file as changed for a single-line modification.
    \item I demonstrate that merging in Git can take exponential time in the number of commits in the history, providing a concrete example illustrating this behavior.
    \item I empirically evaluate the impact of diff algorithms on three-way merges and confirm that the \texttt{histogram} and \texttt{patience} algorithms have advantages when merging real-world commits.
    \item I uncover multiple unexpected behaviors in Git's merging process, including that merges and rebases are not commutative, merges can output conflicts even when the same change is made on both branches, and that, depending on the diff algorithm, successful merges can duplicate changes or fail to detect conflicting changes.
    \item I discover two bugs in Git, one of which I have fixed and contributed to the Git project.
\end{itemize}

These findings demonstrate that, while there are many new and interesting approaches found inside Git that could be applied to other systems, there is also a large potential for improvement in this under-researched area of version control systems.

\chapter{Git's data model}

The chapters \ref{section:diffs} and \ref{section:merge} will focus on two core algorithms that are used by Git, computing diffs and merges.
However, it is important to also know the context these algorithms are used in and why they are required.
For this, this section will explain the fundamental data model of Git, both from the user's perspective and how it is represented internally.
Yet, this is not a manual on how to use Git, nor a documentation on all its features.
The focus lies on the core algorithms and procedures, building a simplified model of Git.
The real implementation has to deal with many special cases and implementation details, like subtrees, file permissions, sparse checkouts, character encodings, line endings and many more.

A large part of Git is either only documented as comments in the codebase, as commit messages, or not documented at all.
In these cases, the respective commits (e.g. \commitcite{03f29155}) or file (e.g \sourcecite{xprepare.c}{427}) will be referenced for further information.
The references use version 2.50 of Git, released on June 6, 2025 \commitcite{16bd9f2}.

\section{Using Git}
The workflow of using Git is well documented in multiple manuals.
The book \textit{Pro Git} \cite{chaconProGit2014} can be seen as a de-facto user manual of the official Git command line interface (CLI) and describes almost all user facing operations.
While Git is primarily used through its CLI, it is also often tightly integrated into other software using a library called libGit2 \cite{Libgit22025}.
This work primarily references the implementations in the official CLI, but the general model is equivalent using for example libGit2.

When the user starts tracking changes with Git, they create a \textit{repository}.
Only within a repository Git can track versions of files.

\paragraph{Commits.}
The most important concept of the versioning system of Git is the \textit{commit}.
A commit is a snapshot of the entire repository at some version \cite[p. 28]{chaconProGit2014}.
It is important to think of a commit as a snapshot of the repository and not as a set of changes, even though the CLI often shows commits as changes.
To create a commit, Git allows users to modify the files of the repository with any editor (this version is called the \textit{working tree}).
Afterward, the current versions must first be added to the \textit{index} (sometimes also called staging area).
Finally, the file versions in the index can be used to create a commit.
The index can mostly be considered a convenience feature and implementation detail, which makes it less relevant to this analysis.
A commit does not only record the entire state of the repository at some point in time but also contains some metadata.
This usually includes the author, a timestamp, a commit message among other data.
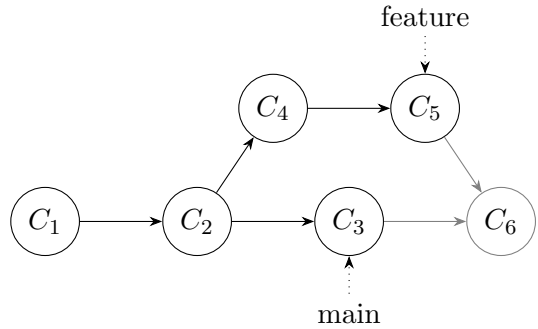
\begin{figure}
    \centering
\begin{tikzpicture}[
    commit/.style={circle, draw=black, minimum size=6mm},
    >=Stealth
  ]

  % Nodes (commits)
  \node[commit] (C1) at (0,0) {$C_1$};
  \node[commit] (C2) at (2,0) {$C_2$};
  \node[commit] (C3) at (4,0) {$C_3$}; % Master continues
  \node[commit] (F1) at (3,1.5) {$C_4$}; % Feature branch
  \node[commit] (F2) at (5,1.5) {$C_5$}; % Feature continues
  \node[commit, draw=black!50] (C6) at (6, 0) {$C_6$}; % Merge commit

  % Edges (arrows showing commit progression)
  \draw[->] (C1) -- (C2);
  \draw[->] (C2) -- (C3);
  \draw[->] (C2) -- (F1);
  \draw[->] (F1) -- (F2);
  \draw[->, draw=black!50] (F2) -- (C6);
  \draw[->, draw=black!50] (C3) -- (C6);

  % Branch labels
  \node[below=.5 of C3] (m) {main};
  \node[above=.5 of F2] (f) {feature};  
  \draw[->,dotted]  (m) --(C3);
  \draw[->, dotted] (f) -- (F2);
\end{tikzpicture}
    \caption{Example of a commit graph.
        The example contains five commits and two branches, \textit{main} and \textit{features}, which have diverged.
        If they are merged, the new commit $C_6$ gets created, which contains the changes of both branches.}
    \label{fig:commits}
\end{figure}
Importantly, commits also contain one or more pointers to their \textit{parents}.
A parent is usually the version of the repository that was modified to reach the state of the commit.
In general, a commit can have arbitrarily many parents, including no parents \cite{GitcheckoutDocumentation}.
Through these parent pointers, the commits of a repository always form a directed acyclic graph (DAG).
An example of this DAG can be seen in \Cref{fig:commits}.

\paragraph{Branches}
While the repository versions are stored in commits, users typically act on \textit{branches}.
A branch is typically thought of as ``a line of development, associated with a specific change'' \cite[p. 160]{lasterProfessionalGit2017}.
In Git's branching model, a branch is simply a named snapshot of the repository, implemented as a pointer to a specific commit.
Users can create new branches, and create commits on any branch that is available.
This way, the branches can \textit{diverge}, meaning that they both contain different commits.
If the user wants to combine both changes, branches can be \textit{merged}.
This will create a merge commit: a new snapshot of the repository that contains the changes from both branches that have been merged.
In \Cref{fig:commits}, the commit $C_6$ would be created when merging the main and feature branches.
Besides a simple merge, Git offers many tools that operate on the DAG of commits.
For example, it is possible to replay the changes of a branch on top of a different base version (rebase a branch) \cite[p. 199]{lasterProfessionalGit2017} or include only the changes of a single commit in another branch (cherry-picking) \cite[p. 203]{lasterProfessionalGit2017}.

\paragraph{Remotes}
To enable collaboration, Git uses a unique decentralized system for synchronizing changes.
In a typical workflow, each user's local repository contains the entire version history of its contents.
Synchronizing changes done by multiple users is accomplished through \textit{remotes}.
Remotes are ``versions of your project that are hosted on the Internet'' \cite[p. 50]{chaconProGit2014}.
Using the push and pull commands users can send their local commits to a remote and download any commits that the remote has into their local repository.
During normal use, this only extends the directed acyclic graph, adding commits but never removing or changing them.
In addition to commits, remotes are also used to synchronize the branches of the repository.
The decentralized architecture allows the use of multiple remotes to exchange specific commits with, even though this is uncommon in practice.

\section{Content addressable storage}
While the above describes how Git is used on a high level, an important part of its design is the content addressable storage system it uses internally.
A repository is always a folder, which contains a \texttt{.git} subfolder.
This \texttt{.git} folder stores the entire version history and metadata.

A commit -- as well as most other data stored by Git -- is stored in its content-addressable object storage and uniquely identified by its hash (object id, \sourcecite{object.h}{158}).
Fundamentally, a commit is a text file that contains the timestamp, message, list of its parent commits, as well as a pointer to a snapshot of the repository, called a \textit{tree} \sourcecite{commit.h}{26}.
A tree does not have to store the entire repository for each commit.
It contains a list of pointers to either \textit{blobs} (raw files) or other trees (subdirectories), together with their paths and file-system metadata (mode) \sourcecite{tree-walk.c}{16}.
Trees and blobs are also stored in the content-addressable object storage and referred to by their hash.
This is the key to how Git is able to store entire snapshots of the repository for each commit:
If a file stays the same between two commits, both trees can use the pointer (hash) to this version, without needing to store the file contents another time.
This applies to trees as well, if an entire subdirectory does not change between two versions.

The content addressable storage system ensures both the immutability of commits (since any change in their content results in a different commit, as identified by its hash) and the acyclicity of the commit graph (every commit contains the SHA-1 hash of its parent, which cannot be changed later to introduce a cycle).
Branches are simply stored as files in the \texttt{.git} folder which contain the hash of the commit they currently refer to.

\section{Model of a Repository}

For the purpose of this work, we define a simplified description of this system.
The goal of this is to preserve most theoretical properties relevant for diffs and merges and ignoring implementation details and optimizations.
A repository $R$ is modeled as a set of commits $C \in R$.
Since git does not track file identities natively, a file $F$ refers to a single version of a file as it is part of a commit.
A single file $F$ is a list of $n$ atoms $a_0, \ldots, a_n$.
During diffing and merging, Git treats lines and the fundamental unit of the file and only operates on entire lines.
Most algorithms only consider whether two lines are identical, ignoring the specific content (some heuristics such as the ones described in \Cref{section:prepostprocessing} are an exception).
However, it can be configured to operate on individual words, separated by spaces, instead \cite[p. 116]{lasterProfessionalGit2017}.
To simplify the language, this thesis refers to the atoms as lines.
While branches are an integral part of Git's usage in practice, the examples in this work will most often simply refer to the commits instead of using branch names.

\section{Finding Differences}
An important part of the system is that the user can use it to find how the tracked content has changed.
Because only snapshots are stored, Git does not know how the user changed the file between two snapshots; the differences have to be computed later if needed.
To compare two versions of a file, Git has multiple algorithms available which are discussed in detail in \Cref{section:diffs}.
The diffing algorithms only compute the difference between two versions of a single file.
However, Git is able to compute differences between entire states of the repository (between two trees).
Git uses a heuristic to detect when files are renamed and copied \sourcecite{diffcore-rename.c}{1037}.
The core of this heuristic consists of computing a similarity score for the content of all pairs of files which have changed in a diff.
If this similarity score is high, this pair is considered a rename or copy, even though it might---in addition to being renamed---have minor changes to its contents \cite{newrenOptimizingGitsMerge2022}.

\section{Merging Branches}
\label{section:datastructure_merge}

% \begin{figure*}
%     \centering
%     \begin{tikzpicture}[
%             commit/.style={circle, draw=black, minimum size=6mm},
%             >=Stealth
%         ]

%         \node[commit] (c1) at (0,0.75) {$C_1$};
%         \node[commit] (c2) at (2,0.75) {$C_2$};
%         \node[commit] (c3) at (4,0) {$C_3$};
%         \node[commit] (c4) at (6,0) {$C_4$};
%         \node[commit] (c5) at (4,1.5) {$C_5$};
%         \node[commit] (c6) at (6,1.5) {$C_6$};
%         \node[commit] (c7) at (8,0.75) {$C_7$};

%         \draw[->] (-1,0.75) -- (c1);
%         \draw[->] (c1) -- (c2);
%         \draw[->] (c2) -- (c3);
%         \draw[->] (c2) -- (c5);
%         \draw[->] (c5) -- (c6);
%         \draw[->] (c6) -- (c7);
%         \draw[->] (c3) -- (c4);
%         \draw[->] (c4) -- (c7);

%         \draw[dotted, rounded corners] (3.25,0.8) rectangle (6.75,2.2);
%         \draw[dotted, rounded corners] (3.25,-0.7) rectangle (6.75,0.7);

%     \end{tikzpicture}
%     \caption{A simple merge of $C_6$ and $C_7$. Their lowest common ancestor is $C_2$, used as a base when merging.
%         The merge can be thought of as applying the changes of both the upper and lower branch to the ancestor $C_2$.}
%     \label{fig:simple_merge}
% \end{figure*}
A merge is used when the repository contains two branches, each with their own changes.
With it, a new version can be created which incorporates the changes of both branches.
While two branches (or rather, their commits) are being merged, a third version plays a major role in the process: the version that both branches are based on.
In the DAG of commits, this is the lowest common ancestor of the two commits being merged.
This simple case for a merge is shown in \Cref{fig:commits}.
The procedure of the merge can be thought of as applying the changes of both branches to the ancestor, rather than directly combining the two commits.

If both branches modified the repository in different areas, it is often possible to apply the changes of both branches independently and thus finish the merge automatically.
However, if the two branches touch similar areas of the repository or even change the same parts, they might \textit{conflict}.
This can happen due to renames which interfere with each other (for example two branches rename the same file, but to a different name) or due to the same file being modified.
In the latter case, Git uses the \textit{three-way merge} or \textit{diff3} algorithm to apply the two changes to the base version.
\Cref{section:merge} analyzes this algorithm in detail.
As with diffs, Git runs a heuristic to find files which might be renamed or copied to identify which files correspond to each other and might require a three-way-merge \cite{newrenOptimizingGitsMerge2022}.
This might also result in conflicts (e.g. conflicts in filenames or permissions), which Git will output and require the user to resolve manually before continuing.

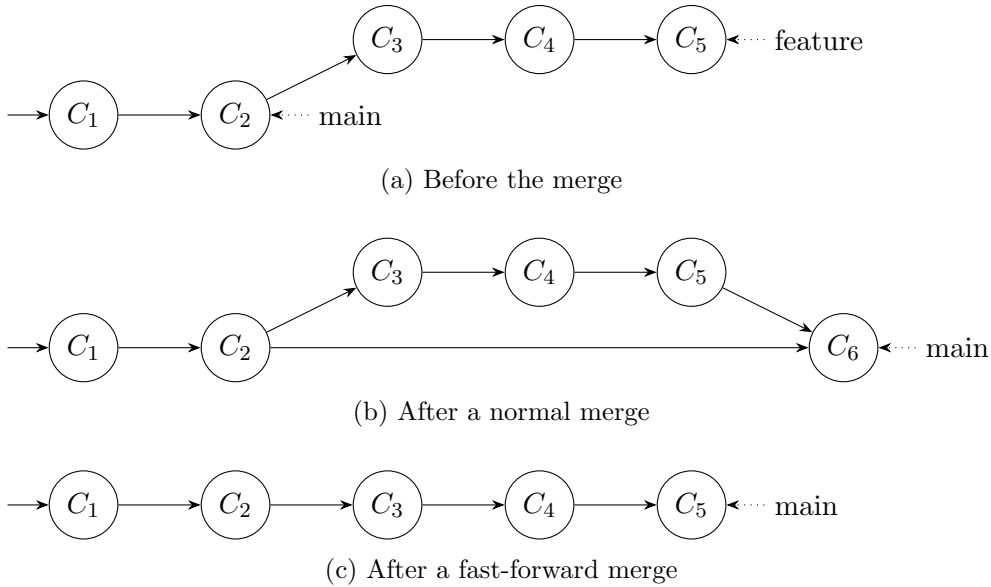
\begin{figure}
    \begin{subfigure}{\textwidth}
        \centering
        \begin{tikzpicture}[
    commit/.style={circle, draw=black, minimum size=6mm},
    >=Stealth
  ]
            \node [commit] (C1) at (0,0)  {$C_1$};
            \node [commit] (C2) at (2, 0) {$C_2$};
            \node [commit] (C3) at (4, 1) {$C_3$};
            \node [commit] (F1) at (6, 1) {$C_4$};
            \node [commit] (F2) at (8, 1) {$C_5$};

            \draw[->] (C1) -- (C2);
            \draw[->] (C2) -- (C3);
            \draw[->] (C3) -- (F1);
            \draw[->] (F1) -- (F2);
            \draw[->] (-1, 0) -- (C1);

            \node[right=.5 of C2] (m) {main};
            \draw[->,dotted] (m) -- (C2);
            \node[right=.5 of F2] (f) {feature};
            \draw[->, dotted] (f) -- (F2);
            \node (phantom) at (11.5,0) {\phantom{main}};
        \end{tikzpicture}   
        \caption{Before the merge}  
    \end{subfigure}
    \begin{subfigure}{\textwidth}
        \centering
        \vspace{.5 cm}
        \begin{tikzpicture}[
    commit/.style={circle, draw=black, minimum size=6mm},
    >=Stealth
  ]
            \node [commit] (C1) at (0,0)  {$C_1$};
            \node [commit] (C2) at (2, 0) {$C_2$};
            \node [commit] (C3) at (4, 1) {$C_3$};
            \node [commit] (F1) at (6, 1) {$C_4$};
            \node [commit] (F2) at (8, 1) {$C_5$};

            \draw[->] (C1) -- (C2);
            \draw[->] (C2) -- (C3);
            \draw[->] (C3) -- (F1);
            \draw[->] (F1) -- (F2);
            \draw[->] (-1, 0) -- (C1);

            \node[commit] (C6) at (10, 0) {$C_6$};
            \draw[->] (F2) -- (C6);
            \draw[->] (C2) -- (C6);

            \node (m) at (11.5,0) {main};
            \draw[->,dotted] (m) -- (C6);
        \end{tikzpicture}  
        \caption{After a normal merge}   
    \end{subfigure}
    \vspace{.5 cm}
        \begin{subfigure}{\textwidth}
                \centering

            \vspace{.5cm}
        \begin{tikzpicture}[
    commit/.style={circle, draw=black, minimum size=6mm},
    >=Stealth
  ]
            \node [commit] (C1) at (0,0)  {$C_1$};
            \node [commit] (C2) at (2, 0) {$C_2$};
            \node [commit] (C3) at (4, 0) {$C_3$};
            \node [commit] (F1) at (6, 0) {$C_4$};
            \node [commit] (F2) at (8, 0) {$C_5$};

            \draw[->] (C1) -- (C2);
            \draw[->] (C2) -- (C3);
            \draw[->] (C3) -- (F1);
            \draw[->] (F1) -- (F2);
            \draw[->] (-1, 0) -- (C1);

            \node[right=.5 of F2] (m) {main};
            \draw[->,dotted] (m) -- (F2);

            \node (phantom) at (11.5,0) {\phantom{main}};

        \end{tikzpicture}    
        \caption{After a fast-forward merge} 
    \end{subfigure}
    \caption{A fast-forward merge. 
    The feature branch is merged into the main branch ($C_5$ and $C_2$ are merged). Their lowest common ancestor is $C_2$, making a fast-forward merge possible.}
    \label{fig:fast_forward_merge}
\end{figure}

\paragraph{Fast-forward merges}
Before running a merge, Git checks for simpler cases where a full merge might not be necessary.
The most important of these cases is the \textit{fast-forward merge}.
If the lowest common ancestor is one of the commits being merged, which means that one commit is an ancestor of the other, the branch pointer can be moved and no merge is necessary.
This is shown in \Cref{fig:fast_forward_merge}.

\section{Advanced Operations}

\begin{figure*}
    \centering
    \begin{tikzpicture}[
            every node/.style={circle, draw=black, minimum size=7mm, inner sep=0, font=\small},    
            >=Stealth
        ]
        \node [] (C1) at (0,0)  {$C_1$};
        \node [] (C2) at (2, 0) {$C_2$};
        \node [] (CM) at (0, 2) {$C_m$};
        \node [] (C2p) at (2, 2) {$C_2'$};

        \draw[->] (-1, 0) -- (C1);
        \draw[->] (C1) -- (C2);
        \draw[->] (C2) -- (3,0);
        \draw[->] ([xshift=-0.6cm] CM.west) -- (CM);
        \draw[->] (CM) -- (C2p);

        \begin{scope}[xshift=6cm, yshift=1cm]
            \node [] (c1) at (0,0) {$C_1$};
            \node [] (c2) at (1, 1) {$C_2$};
            \node [] (cm) at (1, -1) {$C_m$};
            \node [] (c2p) at (2, 0) {$C_2'$};

            \draw[->] (c1) -- (c2);
            \draw[->] (c1) -- (cm);
            \draw[->] (cm) -- (c2p);
            \draw[->] (c2) -- (c2p);
        \end{scope}
    \end{tikzpicture}
    \caption{Cherry-picking using a three-way merge.
    The commit $C_2$ is cherry-picked onto of $C_m$.
    To produce the cherry-picked commit $C_2'$, the three-way merge of $C_2$ and $C_m$, with $C_1$ (the parent of the cherry-picked commit) as a base is used.
    }
    \label{fig:cherry_pick}
\end{figure*}
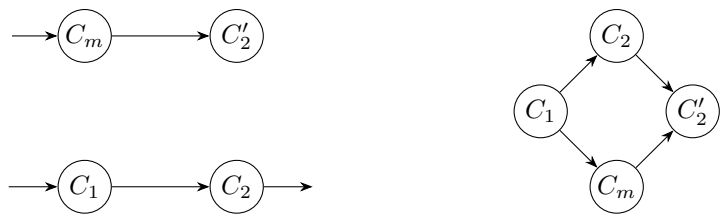

\paragraph{Cherry-picking}
Cherry-picking is the operation of applying the changes of a single commit $A$ in one branch to commit $B$ in a different branch.
Since commits only store a snapshot and not changes, this operation is not as simple as it might seem.
A naive approach to achieve this would be to first compute the changes between $A$ and its parent $P$ and then try to apply these changes to $B$.
This would be an example of \textit{fuzzy patching} since the changes would be applied to a version that might differ from the one the diff was computed from.
However, this is not how Git implements cherry-picking.
Instead, a cherry-pick is implemented as a single merge operation, as shown in \Cref{fig:cherry_pick}.
If a commit $C_2$ should be cherry-picked, $C_1$ is its parent, and the current commit is $C_m$, a new commit $C_2'$ is computed that applies the changes of $C_2$ on top of $C_m$.
To create this commit, a temporary three-way merge is performed, with $C_1$ as the base, $C_2$ as the first parent and $C_m$ as the second parent.
To hide the complexity of this operation, Git sets the only parent of the new commit to $C_m$, resulting in a single revert-commit in a linear history.

\paragraph{Reverting}
An operation that is quite similar to cherry-picking is \textit{reverting}.
This means creating a new commit that has exactly the opposite effect of a previous commit.
Note that one cannot just set the repository state to the old version, since changes that might have been done in the meantime should not be lost (this would be called a \textit{reset} in Git).
Similar to cherry-picking, a revert can be achieved by running a single three-way merge operation, but with the commit and its parent swapped.

When reverting a commit $A$ with a parent $P$ and the current commit is $B$, executing a merge of $P$ and $B$ with the base $A$ will give the desired result.

\paragraph{Rebasing}
With a \textit{rebase}, the point where a branch diverges from another branch can be changed.
This means that all the changes that happened in the branch being rebased (e.g. \texttt{feature}) are replayed on top of another commit.
In many cases this can be used as an alternative to a merge since the resulting branch will have the changes from both branches combined.
The major difference is that no merge commit is needed, simplifying the history.
This operation is implemented as a repeated cherry-pick.
All commits of \texttt{feature} are cherry-picked in order on top of the branch that it is rebased on.

\paragraph{Patches}
Git also has the notion of a \textit{patch}.
A patch can be created with \texttt{git diff} or \texttt{git format-patch} and contains the diff between two trees.
These patch files can be applied using \texttt{git apply}.
While not all workflows make use of patches, the Linux kernel and Git itself are examples of projects which depend strongly on sending patch files via email for collaboration.
When given a simple patch (e.g. one generated by the diff tool of GNU diffutils \cite{GNUDiffutils}), this performs fuzzy-patching and tries to apply the patch based on line-numbers and the given context around diff hunks.
If the file has been modified too much, this fails and the patch has to be applied manually.
However, patches generated using \texttt{git format-patch} usually include hashes of the files that were diffed.
If the patch is applied in the same repository and these versions are available, even applying patches will be done by running a three-way merge.
By using the full merge machinery this will result in a cleaner merge and simpler to resolve conflicts.

\chapter{Diffs}
\label{section:diffs}
Because Git is snapshot-based, computing differences between two versions of a file plays a major role in the system.
Computing differences between two versions of a file becomes important in such a system for two reasons.
The first one is that they are shown to the user.
Creating a good diff helps users understand how their files changed over time, which is a major reason to use a version control system in the first place.
Secondly, diffs are used internally for operations like rebasing, merging, cherry-picking, reverting, and git-blame. % TODO citation needed
Despite this, the diffing algorithms are one of the least well understood parts of the entire system of Git and not well-studied in general.

A basic yet important fact about diffs is that they are not unique.
While one could naively assume that they show what has been changed and there has only been one true change, this is not the case.
For two recorded versions, there are often many different ways of adding and removing lines to get from one version to the other.
Accordingly, multiple different strategies have been added to Git over time \sourcecite{diff.c}{220}.
The default option is called \texttt{myers}.
In addition, Git has the options \texttt{minimal}, \texttt{patience}, and \texttt{histogram}.
These do not only differ in the algorithm used or performance characteristics, but also output different diffs.

\paragraph{History}
In the beginning, Git did not have its own diffing system and instead relied on external tools.
In 2006, the libXDiff library \cite{davidelibenziLibXDiff} was forked at version 0.17 to become Git's diffing system \commitcite{3443546f}.
To trace the history further back, the code of libXDiff is very similar to the original implementation of GNU diffutils \cite{GNUDiffutils} from 1992.
While they are not exactly the same and no clear reference is given, it looks like the text diffing tools of libXDiff were at least inspired by GNU diffutils.

The default \texttt{myers} option uses an implementation of the classic Myers algorithm \cite{myersONDDifferenceAlgorithm1986} that originated from the libXDiff library.
The \texttt{minimal} option was also already part of libXDiff and uses mostly the same code, but disables a heuristic to make sure the resulting diffs are as small as possible.
The other two options, \texttt{patience} and \texttt{histogram} were added later, with \texttt{histogram} created as an improvement upon \texttt{patience}.

\section{Representations}
Before describing the diff algorithms implemented in Git in detail, we need some basic understanding of what a diff even is.
Intuitively, a diff shows the difference between two versions of a file (or two files in general).
The diff algorithm operates on two finite sequences $A = (a_1, a_2, \ldots, a_n)$ and $B = (b_1, b_2, \ldots, b_m)$.
In the following, these will be called the \textit{old} and \textit{new} files for simplicity, even though the algorithm does not require them to be ordered in any way.
By default, $A$ and $B$ will be sequences of entire lines of source code, since Git considers a line as the atomic unit when diffing.
Within this structure, there are multiple ways to represent a diff:
\begin{description}
    \item[Changed lines] The representation used internally when computing a diff are changed lines.
        For each line in the old and new file, a flag is used that represents whether this line was changed.
        A changed line in the old file indicates a deletion, while a changed line in the new file indicates an addition.
        This representation is what is used when displaying a side-by-side diff, highlighting lines in each version.
    \item[Edit scripts] Formally, an \textit{edit script} is a sequence of insertion or deletion instructions that transforms the old file into the new file.
        While usually an instruction can only add or delete a single line, in Git an \textit{edit script} refers to a series of \textit{hunks}.
        Hunks group changes (additions and deletions) that appear at the exact same point in the file together in a single block.
        This format is similar to the one used in the output of \texttt{git diff} (it shows multiple hunks in a single output block when they are close together).
        A hunk can efficiently be represented as a region in the old file and a region in the new file.
        The semantic meaning is that the lines in the old file have to be deleted and replaced by the lines in the new file.
    \item[Matchings] The same diffs can also be thought of as a matching between the lines of the first file and the lines of the second file, matching two lines if and only if the algorithm determined them to be unchanged.
        However, matchings are a more powerful representation than diffs in Git.
        Git does not analyze reordering operations which could in theory be expressed by a matching.
        Nevertheless, for each diff generated by Git, a corresponding matching between the lines of the files can be computed.
    \item[Common subsequences] Since the unchanged lines have to be the same in both files in a correct diff, a diff can be represented as a common subsequence of the two files.
        $C = (c_1, c_2, \ldots, c_k)$ is called a common subsequence of $A$ and $B$ iff there exist indices $1 \leq i_1 < i_2 < \cdots < i_k \leq n$ and $1 \leq j_1 < j_2 < \cdots < j_k \leq m$ where $C\left[l\right] = A\left[i_l\right] = B\left[j_l\right], \,\,\,\forall l \in \{1, 2, \ldots, k\}$.
        Lines that are in $A$ but not in $C$ are displayed as deleted, while lines in $B$ but not in $C$ are displayed as added.
        This way, every valid common subsequence can be used to derive a changed-lines representation of a diff.
\end{description}
Two important observations concerning these representations are:
\begin{enumerate}
    \item Diffs generated from common subsequences are always \textit{correct}, in the sense that applying the related edit script to the old file will result in the new file.
    \item The longer the common subsequence, the shorter the edit script.
\end{enumerate}
It follows that finding a \textit{longest} common subsequence (LCS) will result in the shortest possible diff, which will also always be correct \cite{wagnerStringtoStringCorrectionProblem1974}.
All LCS-based algorithms minimize the length of the edit script.
However, the shortest edit script (a \textit{minimal} diff) is not unique.
An example of this is $A = ab$, $B = ba$---both $C_1 = a$ or $C_2 = b$ are LCS of $A$ and $B$.
Furthermore, often an edit script that is not minimal may better reflect the actual editing process than a minimal diff (see \Cref{fig:histogram_good} for an example).
Defining which diff is best from the user's point of view is therefore difficult. 
It is often subjective and highly dependent on the change and context.
While the Myers and minimal algorithms both try to find diffs based on an LCS, the other two algorithms in Git aim to find subjectively ``better'' diffs by applying heuristics.

\section{Myers and Minimal Diff Algorithms}
Myers describes two variations of his $O((N+M)D)$ (which he simplified to $O(ND)$) algorithm for computing shortest edit scripts ($N$ and $M$ are the lengths of the two sequences to be compared, $D$ is the length of the minimum edit script).
The first is a simple version, which uses bottom-up dynamic programming and $O(D^2)$ space, and the second is a divide-and-conquer solution that only uses $O(N+M)$ space \cite{myersONDDifferenceAlgorithm1986}.
The \texttt{minimal} and \texttt{myers} options in Git both use a common implementation that closely follows the divide-and-conquer version of Myers algorithm.
With the \texttt{myers} option, two additional heuristics are enabled to shortcut the algorithm on very large instances.
Despite its name, these heuristics were not part of the original algorithm by Myers.
The \texttt{minimal} option implements Myers algorithm almost exactly.

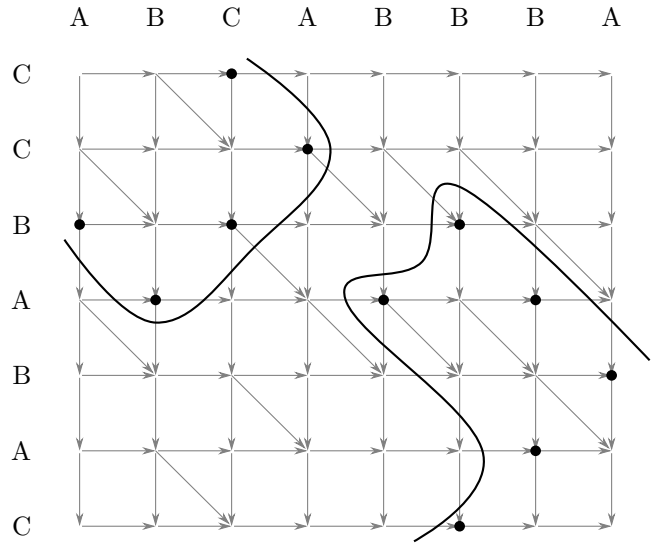
\begin{figure}
    \centering
    \begin{tikzpicture}[
    gridpt/.style={circle,inner sep=0.5pt},
    arrowfield/.style={-{Stealth[length=1.8mm, width=1mm]},draw=gray},
    arrowfieldrev/.style={{Stealth[length=1.8mm, width=1mm]}-,draw=gray},
    dblpath/.style={->,thick},
    top label/.style={font=\small,anchor=south},
    left label/.style={font=\small,anchor=east}
  ]

  % grid size
  \def\NX{7}   % x = 0..7
  \def\NY{6}   % y = 0..6

    % define letter arrays
    \def\colletters{{"A","B","C","A","B","B","B","A"}}
    \def\rowletters{{"C","C","B","A","B","A","C"}}

  % draw all grid‐points and the little 3‐arrow field at each point
  \foreach \x in {0,...,\NX} {
    \foreach \y in {0,...,\NY} {
      % place the dot
      \node[gridpt] (P\x\y) at (\x,-\y) {};

      % right‐pointing arrow (if not on right boundary)
      \ifnum\x<\NX
        \draw[arrowfield] (P\x\y) -- ++(1,0);
      \fi
      % up‐pointing arrow (if not on top row y=0)
      \ifnum\y<\NY
        \draw[arrowfield] (P\x\y) -- ++(0,-1);
      \fi
    % diagonal down-right (if x<NX and y<NY)
    \ifnum\x>0
        \ifnum\y>0
          \pgfmathparse{\colletters[\x]}
          \edef\colletter{\pgfmathresult}
          \pgfmathparse{\rowletters[\y]}
          \edef\rowletter{\pgfmathresult}
          \ifx\colletter\rowletter
            \draw[arrowfieldrev] (P\x\y) -- ++(-1,1);
          \fi
        \fi
    \fi 
    }
  }

  % labels 
    \foreach \x in {0,...,\NX} {
        \pgfmathparse{\colletters[\x]}
        \edef\colletter{\pgfmathresult}
        \node[top label] at (\x,0.5) {\colletter};
    }
    \foreach \y in {0,...,\NY} {
        \pgfmathparse{\rowletters[\y]}
        \edef\rowletter{\pgfmathresult}
        \node[left label] at (-0.5,-\y) {\rowletter};
    }

    % store a list of the points into the "forward" array
    \foreach \x/\y in {2/0, 3/1, 2/2, 1/3, 0/2} {
        \fill (P\x\y) circle (2pt);
    }
    \draw[thick] plot [smooth, tension=0.7] coordinates { (2.2, 0.2) (3.3, -1) (2.3, -2.3) (1, -3.3) (-0.2, -2.2)  };

    \foreach \x/\y in {7/4, 6/3, 5/2, 4/3, 5/6, 6/5} {
        \fill (P\x\y) circle (2pt);
    }

    \draw[thick] plot [smooth, tension=0.7] coordinates { (7.50, -3.8) (5, -1.5) (4.5, -2.5) (3.5, -3) (5.3, -5) (4.4, -6.2)};

\end{tikzpicture}
    \caption{Edit graph of the Myers algorithm (example from Myers \cite{myersONDDifferenceAlgorithm1986}). The highlighted boundaries show the points reachable with a cost of 2.}
    \label{fig:edit-graph}
\end{figure}

The Myers algorithm works roughly as follows:
Like many algorithms based on the dynamic programming technique, Myers algorithm works on a tabular data format, called the \textit{edit graph}.
\Cref{fig:edit-graph} shows the edit graph for finding the longest common subsequence between $A = $ \texttt{ABCABBBA} and $B = $ \texttt{CCBABAC}.
The two sequences are written along the horizontal or vertical edge, respectively.
Each vertex $v_{x,y}$ corresponds to a position in the two sequences, meaning that the first $x$ lines of $A$ and $y$ lines of $B$ have been considered.
Every vertex has a cost $c(v)$, which corresponds to how many lines had to be added/removed to get to this vertex.
It can be computed from the following recurrence relation:
\begin{align}
    c(v_{x,y}) = \begin{cases}
                     0                                    & \text{if } x = y = 0   \\
                     c(v_{x-1,y-1})                       & \text{if } A[x] = B[y] \\
                     1 + \min(c(v_{x-1,y}), c(v_{x,y-1})) & \text{otherwise}
                 \end{cases}
    \label{eq:myers_cost}
\end{align}
The algorithm computes the minimum cost of the last vertex $v_{N, M}$ and then reconstructs the path from there.
This is equivalent to computing a weighted shortest path from $v_{0,0}$ (top-left corner) to $v_{N, M}$ (bottom-right corner).
Every vertex is connected to its bottom and right neighbor, representing that a line has to be removed or added, respectively (with a cost of 1).
The diagonal connection represents that the line does not have to be changed and has a cost of 0---if the two lines are equal.
A path from $v_{0,0}$ to $v_{N, M}$ represents an edit script.

While this principle is the basis for multiple edit-distance-based algorithms, the divide-and-conquer version of Myers algorithm employs a more efficient approach than just searching for the shortest path using an approach like Dijkstra's algorithm.
A breadth-first search is started from both the starting point $v_{0,0}$ and the ending point $v_{N, M}$.
For each search, the frontier can be stored by storing, for each diagonal $d = x - y$, the furthest vertex $v_{x,y}$ that can be reached with a cost of $c$.
During each iteration, \Cref{eq:myers_cost} is applied to compute the vertex reachable with a cost of $c+1$.
When the two searches meet at a vertex $v_{x,y}$ (the \textit{pivot}), it follows by the optimal substructure of the shortest-path problem that this vertex has to be on the shortest path from $v_{0,0}$ to $v_{N, M}$.
The algorithm then records this vertex and recurses again on the path from $v_{0,0}$ to $v_{x,y}$ and from $v_{x,y}$ to $v_{N, M}$.
This way, the algorithm does not have to store back-pointers to reconstruct the path, reducing the space complexity from $O(D^2)$ to $O(N+M)$, as only the current frontier has to be stored.
The original paper by Myers \cite{myersONDDifferenceAlgorithm1986} describes the algorithm in more detail, including how exactly the breadth-first search is implemented, which the implementation in Git closely follows.

\paragraph{Heuristics in Git's \texttt{myers} option}
The \texttt{myers} option in Git does not implement the Myers algorithm exactly, but instead adds two heuristics on top of it.
These are primarily intended to speed up the algorithm in cases where it would otherwise take a long time to run.
They are both disabled when using the \texttt{minimal} option, which implements the Myers algorithm as described above.

The first heuristic is concerned with so-called \textit{snakes}.
A snake is a path that ends with at least 20 continuous diagonal steps (matching equal lines)\sourcecite{xdiffi.c}{28}.
If one or more snakes have been found, and the algorithm has already searched for 256 steps \sourcecite{xdiffi.c}{26} from both directions, the heuristic is activated \sourcecite{xdiffi.c}{152}.
The algorithm will now compute a score for each of the points on the current frontiers as the total progress in each of the sequences minus the distance to the diagonal across the edit graph.
Among the points which are at the end of a snake, the algorithm will choose the one with the highest score and use it as the pivot, even though the searches have not met yet, and it might not be part of a shortest path.
A possible intuition for this logic is that in files that require more than 512 changes, when the algorithm has found a way to edit both sequences such that the next 20 lines line up perfectly, it just assumes that the edits are correct and no longer considers alternatives.

The second heuristic is always activated when the search has been running for at least $\mathop{min}(\mathop{approxSqrt}(N), 256)$ steps \sourcecite{xdiffi.c}{207} ($N$ is the number of lines in the input).
The function $\mathop{approxSqrt}(n)$ is defined as the smallest power of two that is larger than $\sqrt{n}$ : $\mathop{approxSqrt}(n) = 2^{\lceil \log_2(\sqrt{n}) \rceil}$ (this is used as an optimization in the implementation) \sourcecite{xprepare.c}{378}.
In that case, the search is terminated and the point furthest from the starting point (or ending point) is used to pivot on.
This is just a simple shortcut taken to avoid long execution times in files with a very large number of changes.

These two heuristics never result in an incorrect diff, since the algorithm will still find a common subsequence, just not the longest one.
The empirical evaluation in \Cref{section:comparison} analyzes how much longer the diffs get and how much performance is saved compared to using the \texttt{minimal} option, which does not use these heuristics.
The effect of these heuristics is contained to cases where a very large number of changes is necessary and even then, the diffs are only slightly longer.

\section{Patience Algorithm}
The \textit{patience} algorithm available in Git is mostly of historical significance, since it inspired ideas found in the later \textit{histogram} algorithm \cite{pearceEclipseJGitImplement2010}.
It is credited to Bram Cohen, who published an informal description of the algorithm and its motivation on his blog in 2010 \cite{cohenPatienceDiffAdvantages2010}.
The rationale behind it is that lines that are \textit{unique}, in the sense that they appear only a single time within each file, are more important than the other lines and should be preserved rather than added or removed \cite{habouzitLibxdiffPatienceDiff2008}.
In typical source code, unique lines could be for example function signatures or lines containing logic that is not duplicated within the file.
Non-unique lines are usually blank, or lines containing single braces, start/end comment markers or common function calls.
Thus, the idea is to, if possible, mark the non-unique lines as changed to preserve the unique lines, as they are more important to the overall structure of the program.

% \documentclass{standalone}

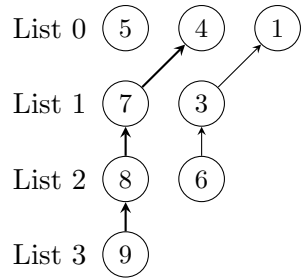
\begin{figure}
    \centering
    % \documentclass[tikz,border=5pt]{standalone}
% \begin{document}
\begin{tikzpicture}[
    entry/.style={circle,draw,minimum size=6mm,inner sep=0pt,font=\small},
    arrow/.style={->,>=stealth,thick}
  ]

\node[entry] (p5) at (0,0) {5};
\node[entry] (p4) at (1,0) {4};
\node[entry] (p1) at (2,0) {1};

\node[entry] (p7) at (0,-1) {7};
\node[entry] (p3) at (1,-1) {3};

\node[entry] (p8) at (0,-2) {8};
\node[entry] (p6) at (1,-2) {6};

\node[entry] (p9) at (0,-3) {9};

  \draw[arrow] (p7) -> (p4);
    \draw[arrow] (p8) -> (p7);
    \draw[arrow] (p9) -> (p8);

    \draw[arrow, thin] (p6) -> (p3);
    \draw[arrow, thin] (p3) -> (p1);
  
\node (s1) at (-1, 0) {List 0};
\node (s2) at (-1, -1) {List 1};
\node (s3) at (-1, -2) {List 2};
\node (s4) at (-1, -3) {List 3};
\end{tikzpicture}
% \end{document}
    \caption{An example of patience sorting for the input sequence \texttt{5, 4, 7, 8, 1, 3, 9, 6} (from \cite{mallowsProblem622Patience1962}).
        The algorithm inserts the numbers in order, always at the end of the first list which has a last value greater or equal to the current number, recording the current last value of the previous list as a back pointer.
        From these pointers, a longest increasing subsequence can be reconstructed.
        In the given example, a longest increasing subsequence is \texttt{4, 7, 8, 9}.}
    \label{fig:patience_lcs}
\end{figure}

% \usepackage{algorithm2e}
% \usepackage{amsmath}
% \usepackage{amstext}
% \begin{document}
\begin{algorithm}
\caption{Pseudocode description of the patience diff algorithm.}
\label{alg:patience_diff}
\SetAlgoLined

\SetKwFunction{PatienceDiff}{PatienceDiff}
\SetKwFunction{WalkLcs}{WalkLcs}
\SetKwFunction{FindMatchingUniqueLines}{FindMatchingUniqueLines}
\SetKwFunction{BinarySearch}{BinarySearch}
\SetKwFunction{FindLcs}{FindLcs}

\SetKwData{Lcs}{lcs}

\SetKwData{SeqA}{seq1}
\SetKwData{SeqB}{seq2}
\SetKwData{Entry}{entry}
\SetKwData{Unique}{unique}

\SetKwProg{Fn}{Function}{:}{}

\Fn{\PatienceDiff{\SeqA, \SeqB}}{
   \If{$\left|\SeqA\right| = 0 \lor \left|\SeqB\right| = 0$}{
       Mark entire \SeqA or \SeqB as changed\\
       \Return{}
   } 

    \Unique $\leftarrow$ \FindMatchingUniqueLines(\SeqA, \SeqB)\\

   \Lcs $\leftarrow$ \FindLcs(\Unique)\\

   \uIf{$\neg \Lcs$} {
        fall back to other diff algorithm\\
   } 
   \Else {
        recursively call \PatienceDiff on the segments between \Lcs
   }
}
\vspace{0.5cm}

\Fn{\FindMatchingUniqueLines{\SeqA, \SeqB}}{
    \SetKwData{Line}{line}
    $\Unique \leftarrow \left[\right]$ \\

    \For{$\Line \leftarrow 0$ \KwTo $|\SeqA| - 1$}{
        \If{\Line is unique in \SeqA $\land$ \Line is unique in \SeqB}{
            add a record with both indices to \Unique\\
        }
    }
    \Return{\Unique}\\
}
\vspace*{0.5cm}
\Fn{\FindLcs{\Unique}}{
    \SetKwData{Sequence}{sequences}
\tcc{This is the patience sorting algorithm for finding a longest increasing subsequence.}
    $\Sequence \leftarrow \left[\right]$\\
    \For {$\Entry \in \Unique$} {
        \SetKwData{I}{i}

        $\I \leftarrow$ last index in \Sequence where $\text{last}(\Sequence\left[\I\right]).\mathit{posB} < \Entry.\mathit{posB}$\\
        \Entry.\textit{previous} $\leftarrow$ \text{last}($\Sequence\left[\I\right]$) \\
        $\I \leftarrow \I+ 1$\\
        \uIf{$\I < \left| \Sequence \right|$}{
            $\Sequence\left[\I\right]$.push(\Entry)\\
        } \Else {
            \Sequence.push($\left[\Entry\right]$)\\
        }
    }

    \SetKwData{Lcs}{lcs}
    $\Lcs \leftarrow$ reconstruct LCS by walking back \textit{previous} pointers from the last element of the last list in \Sequence\\

    \Return {\Lcs}\\
}
\vspace{0.5cm}
\end{algorithm}
% \end{document}
\Cref{alg:patience_diff} shows pseudocode for a simplified version of the patience algorithm as implemented in Git.
The overall principle of the algorithm is that, among the lines that are common among to two files and unique, it computes the longest common subsequence.
Then, it recurses between the lines that are part of this common subsequence.
To do this, it first builds up a list of all such lines, recording their position in both files, ordered by their appearance in the first file.
The actual implementation uses a custom hash map to do this efficiently \sourcecite{xpatience.c}{142}.
The \texttt{FindLcs} function finds the longest common subsequence among these lines by using an algorithm for the longest \textit{increasing} subsequence.
This is based on the fact that, when $A$ is a permutation of $B$ ($\forall x: x\in A \iff x\in B$) and all lines are unique~($\forall x,y: x = y \lor A[x] \neq A[y] \land B[x] \neq B[y]$), the longest common subsequence can be seen as the longest increasing subsequence of the permutation (with the permutation being the list of the indices of the lines in $B$).
Since the lines are already stored in their order of appearance in the first file, finding the longest increasing subsequence of the positions in the second file will yield a longest common subsequence among the unique lines.
In order to find the longest increasing subsequence, \textit{patience sorting} \cite{mallowsProblem622Patience1962,bespamyatnikhEnumeratingLongestIncreasing2000} is used, which gives the algorithm its name.
\Cref{fig:patience_lcs} illustrates an example of how patience sorting finds a longest increasing subsequence.
Finally, the algorithm splits the file along the found longest common subsequence of unique lines and recurses in between them.
If no common unique lines are present in the two sequences, the algorithm falls back to the Myers algorithm.

The patience algorithm remains available in Git today and is also part of a few other version control systems, like Mercurial and Bazaar.
It is commonly noted that it runs slower than the other algorithms, which seems to originate from a measurement done after histogram was added \commitcite{85551232}.
In \Cref{section:comparison}, a detailed comparison of the algorithms follows, which also empirically analyzes its running time.
It also contains examples of where the patience algorithm produces nicer looking diffs.

\section{Histogram Algorithm}
The histogram algorithm was written as a version of the patience algorithm for JGit, by Shawn O. Pearce \cite{pearceEclipseJGitImplement2010}.
In 2011, the original Java implementation was closely translated to C and included in Git \commitsourcecite{8c912eea}{xhistogram.c}{1}.
When introduced to Git, the primary reason mentioned for its inclusion was the superior performance compared to the Myers algorithm and the patience diff.
The commit message in JGit \cite{pearceEclipseJGitImplement2010} explains the idea of the algorithm and seems to be the only public documentation of the algorithm by Shawn O. Pearce:

\begin{quote}
    HistogramDiff is an alternative implementation of patience diff,
    performing a search over all matching locations and picking the
    longest common subsequence that has the lowest occurrence count.
    If there are unique common elements, its behavior is identical to
    that of patience diff.
\end{quote}

There is no further documentation of the algorithm included, and it is severely underdocumented:
the source code itself contains almost no comments and uses a very low-level style of C (and Java), increasing the difficulty of understanding the algorithm.
Nevertheless, it is currently the default algorithm when merging \cite{GitMergestrategiesDocumentation} and often recommended as the most modern diffing algorithm available \cite{nugrohoHowDifferentAre2020a,linustorvaldsLinux64rc12023}.

\paragraph{Related work}
There have been multiple attempts at documenting or even reimplementing the histogram algorithm.
In a blog post explaining multiple diff algorithms, Tiark Rompf \cite{tiarkrompfHistogramDiffAlgorithm} gives some pseudocode and an example implementation in JavaScript.
In this version, the algorithm always splits the entire file on the first least-frequent line and recurses before and after this line.
Nugroho et al. \cite{nugrohoHowDifferentAre2020a} conducted an empirical evaluation of the diff algorithms for computing software engineering metrics.
They give an informal description and example of the algorithm which includes searching both files for infrequent lines and selecting an LCS from that.
In the version control system Jujutsu \cite{jjcontributorsJjvcsJj2025}, a similar algorithm to histogram was recently implemented \cite{martinvonzweigbergkJJVCSDiff}.
This implementation is well-documented and closely follows the patience algorithm: It computes the LCS among the least-frequent (instead of unique) lines, splitting the file and recursing on the parts.
All these efforts show slightly different algorithms and do not agree on how it works exactly.
The implementation in Jujutsu extends the patience algorithm for the case where there are no unique lines in the file.
This is what the last sentence of the above description seems to imply.
The code of Tiark Rompf does not even contain an algorithm for finding a longest common subsequence, focusing more on the first sentence of the above description.
Ray Gardner \cite{raygardnerHowHistogramDiff2025} published a pseudocode version of the algorithm that has been derived from the implementation and is much closer to what the algorithm actually does.

\paragraph{Algorithm description}
In fact, the histogram algorithm is more ``inspired by'' the patience algorithm than an extension of it.
It works significantly differently, does not include the patience sort procedure, and is not even based on longest common subsequences.
To give the following description of the algorithm, I have reverse engineered the C implementation in Git \sourcecite{xhistogram.c}{}.
To do this, I have first translated it to Python, carefully abstracting the custom hash-map implementation that is part of the C implementation, and finally translated it to pseudocode.
\Cref{alg:histogram_diff_1} and \Cref{alg:histogram_diff_2} show this derived pseudocode.

\begin{algorithm}
    \small
    \caption{Pseudocode of the histogram diff algorithm as found in Git 2.50. The implementation makes use of a
        custom hash map implementation for the occurrences map, which has been omitted here. If a bucket of that
        implementation is filled with more than 64 elements, the algorithm fails and falls back to a different diff algorithm.}
    \label{alg:histogram_diff_1}
    \SetAlgoLined

    \SetKwFunction{ScanA}{ScanA}
    \SetKwFunction{TryLcs}{TryLcs}
    \SetKwFunction{FindLcs}{FindLcs}
    \SetKwFunction{HistogramDiff}{HistogramDiff}

    \SetKwData{Index}{index}
    \SetKwData{Record}{record}
    \SetKwData{Lcs}{lcs}
    \SetKwData{Split}{split}
    \SetKwData{SeqA}{seq1}
    \SetKwData{SeqB}{seq2}

    \SetKwProg{Fn}{Function}{:}{}
    \Fn{\ScanA{\SeqA}}{
    \tcc{Indices where a line can be found}
    $\Index.\mathit{ocurrences} \leftarrow \left\{\right\}$ \\

    \For{\textit{line} $\leftarrow$ $|\SeqA| - 1$ \KwTo $0$}{
    \uIf{$\Index.\mathit{ocurrences}\left[\SeqA\left[\mathit{line} \right]\right]$}{
        $\Index.\mathit{ocurrences}\left[\SeqA\left[\mathit{line} \right]\right].\texttt{push\_front}(\textit{line})$\\
    }
    \Else{
        $\Index.\mathit{ocurrences}\left[\SeqA\left[\mathit{line} \right]\right] \leftarrow \left[ \mathit{line} \right]$ \\
    }
    }
    \Return{\Index}\\
    }
    \vspace{0.4cm}

    \Fn{\FindLcs{\SeqA, \SeqB}}{
        \Index $\leftarrow$ \ScanA(\textit{\SeqB})\\

        \textit{seq2Index} $\leftarrow 0$\\
        \Lcs $\leftarrow$ \textbf{new} LCS(start1: 0, end1: 0, start2: 0, end2: 0)\\
        \While{seq2Index $<|\SeqB|$}{
            \textit{seq2Index}, \Lcs $\leftarrow$ \TryLcs(\textit{seq2Index}, \Index)\\
        }
        \If {\Index.has\_common $\land$ \Index.lowestRecordCount $>$ 64}{
            \Return{\textbf{None}}\\
        }
        \Return{\Lcs}\\
    }
    \vspace{0.4cm}

    \Fn{\HistogramDiff{\SeqA, \SeqB}}{
        \If{$\left|\SeqA\right| = 0 \land \left|\SeqB\right| = 0$}{
            \Return{}
        }

        \uIf{$\left|\SeqA\right| = 0$}{
            mark $\text{\SeqA}$ as changed\\
            \Return{}
        }
        \ElseIf{$\left|\SeqB\right| = 0$}{
            mark $\text{\SeqB}$ as changed\\
            \Return{}
        }

        \Split $\leftarrow $ \FindLcs(\SeqA, \SeqB)\\
        \uIf{$\neg$ \Split}{
            fallback to other diff algorithm \\
        } \uElseIf{\Split.start1 = 0 $\land$ \Split.start2 = 0}{
            mark \SeqA and \SeqB as changed\\
        } \Else{
            $\HistogramDiff(\SeqA\left[:\Split.\mathit{start1}\right], \SeqB\left[:\Split.\mathit{start2}\right])$\\
            $\HistogramDiff(\SeqA\left[\Split.\mathit{end1} + 1:\right], \SeqB\left[\Split.\mathit{end2} + 1:\right])$\\
        }
    }
    \vspace{0.4cm}
\end{algorithm}

\begin{algorithm}
    \caption{Pseudocode of the \texttt{TryLcs} function of the histogram diff algorithm.
        It iterates over multiple potential regions in the first file that could be matched with the given position in the second file, selecting one based on length or frequency of the least frequent contained line.}
    \label{alg:histogram_diff_2}
        \SetAlgoLined
    \small
    \SetKwFunction{ScanA}{ScanA}
    \SetKwFunction{TryLcs}{TryLcs}
    \SetKwFunction{FindLcs}{FindLcs}
    \SetKwFunction{HistogramDiff}{HistogramDiff}

    \SetKwData{Index}{index}
    \SetKwData{Record}{record}
    \SetKwData{Lcs}{lcs}
    \SetKwData{Split}{split}
    \SetKwData{SeqA}{seq1}
    \SetKwData{SeqB}{seq2}

    \SetKwProg{Fn}{Function}{:}{}

    \Fn{\TryLcs{seq2Index, \Index, \Lcs}}{
        \tcc{Find LCS at current position in seq2}
        \textit{bNext} $\leftarrow$ \textit{seq2Index} + 1\\
        \textit{count} $\leftarrow$ $\left|\Index.\mathit{ocurrences}\left[seq2\left[\mathit{seq2Index}\right]\right]\right|$\\

        \If{$\left|\Index.\mathit{ocurrences}\left[seq2\left[\mathit{seq2Index}\right]\right]\right| = 0$ } {
            \Return{\textit{bNext}, \Lcs}\\
        }

        \Index.has\_common $\leftarrow$ 1\\

        \If{\textit{count} $>$ \Index.lowestRecordCount}{
            \Return{\textit{bNext}, \Lcs}\\
        }
        \textit{endSeqA} $\leftarrow$ 0\\
        \For{\textit{seqAPointer} \textbf{\textrm{in}} $\Index.\mathit{ocurrences}\left[seq2\left[\textit{seq2Index}\right]\right]$}{
            \If{\textit{seqAPointer} $<$ \textit{endSeqA}}{
                \textbf{continue}\\
            }

            \textit{beginSeqB} $\leftarrow$ \textit{seq2Index}\\
            \textit{endSeqA} $\leftarrow$ \textit{seqAPointer}\\
            \textit{endSeqB} $\leftarrow$ \textit{beginSeqB}\\

            \tcc{Extend region backwards}
            \While{\textit{seqAPointer} $>$ 0 \textbf{and} \textit{beginSeqB} $>$ 0 \textbf{and} $\mathit{seq1}\left[\textit{seqAPointer} - 1\right] = \mathit{seq2}\left[\textit{beginSeqB} - 1\right]$}{
                \textit{seqAPointer} $\leftarrow$ \textit{seqAPointer} - 1\\
                \textit{beginSeqB} $\leftarrow$ \textit{beginSeqB} - 1\\
            }

            \tcc{Extend region forwards}
            \While{\textit{endSeqA} $<$ $|$seq1$|$ - 1 \textbf{and} \textit{endSeqB} $<$ $|$seq2$|$ - 1 \textbf{and} $\mathit{seq1}\left[\textit{endSeqA} + 1\right] = \mathit{seq2}\left[\textit{endSeqB} + 1\right]$}{
                \textit{endSeqA} $\leftarrow$ \textit{endSeqA} + 1\\
                \textit{endSeqB} $\leftarrow$ \textit{endSeqB} + 1\\
            }

            \tcc{Find minimum frequency in the region}
            recordCount $\leftarrow$ $\min(\left[\left|\Index.\mathit{ocurrences}\left[\SeqA\left[line\right]\right]\right| \textbf{for}\,\, \mathit{line}\,\, \textbf{in}\,\, \textit{seqAPointer}..\textit{endSeqA}\right])$\\

            \If{bNext $\leq$ \textit{endSeqB}}{
                \textit{bNext} $\leftarrow$ \textit{endSeqB} + 1\\
            }

            \tcc{Update LCS if better one found}
            \If{\Lcs.end1 - \Lcs.begin1 $<$ \textit{endSeqA} - \textit{seqAPointer} \textbf{or} recordCount $<$ \Index.\textrm{lowestRecordCount}}{
                \Lcs.begin1 $\leftarrow$ \textit{seqAPointer}\\
                \Lcs.begin2 $\leftarrow$ \textit{beginSeqB}\\
                \Lcs.end1 $\leftarrow$ \textit{endSeqA}\\
                \Lcs.end2 $\leftarrow$ \textit{endSeqB}\\
                \Index.lowestRecordCount $\leftarrow$ recordCount\\
            }
        }

        \Return{bNext, \Lcs}\\
    }
    \vspace{0.3cm}
\end{algorithm}
The overall structure of the algorithm is recursive.
First, it searches for a common contiguous sequence of lines (\textit{region}) that can be found in both files.
Then, it splits the file into two parts, before and after the region and calls itself recursively on these two parts.
The recursion ends when one of the files is empty or both files do not have any lines in common anymore, marking the entire files as changed.
The core of the algorithm is found in the procedure to select this common region (unfortunately named \texttt{findLCS} even though the region it finds is not an LCS) \sourcecite{xhistogram.c}{249}.
The function \texttt{scanA} \sourcecite{xhistogram.c}{103} builds up a hash map of where and how often a given line appears in the first file.
Next, the algorithm conducts a search through the second file, starting from the first line (\texttt{tryLCS} \sourcecite{xhistogram.c}{156}).
For the current line in the second file, it iterates over all the occurrences of this line in the first file and tries to find a maximum size region there:
As long as the lines above and below the original matching lines are equal, the region is expanded in both files to find a \textit{maximum region}.
The search continues in the second file with the first line which was not part of the previous maximum region.
To select which of these maximum regions to take as a split point, the algorithm keeps track of the \textit{lowest record count}.
This is, for all the lines in the region, the frequency of the least frequent line.
This frequency is only counted in the first file.
If a region is either larger or has a lower \textit{lowest record count} than the current best, it is remembered as the new best region.
When the algorithm reaches the end of the second file, the current best region is used as a split point.

\paragraph{Relation to the patience algorithm}
It is clear that the way \texttt{tryLCS} works, it does not return a longest common subsequence in the traditional sense.
This seems to be the largest difference to the patience algorithm and also the source of confusion about how it works.
In the histogram algorithm and the commit message describing it, the term longest common subsequence refers to a region rather than a subsequence.
Typically, a subsequence $S'$ of a sequence $S = (s_1, s_2, \ldots, s_n)$ is defined as a sequence $S' = (s_{i_1}, s_{i_2}, \ldots, s_{i_k})$ where $1 \leq i_1 < i_2 < \cdots < i_k \leq n$.
In other words, $S'$ can be obtained by deleting some (possibly zero) elements from $S$ while preserving the relative order of the remaining elements.
On the other hand, a region $r$ of a sequence $s$ is defined as a contiguous subsequence $r = (s_i, s_{i+1}, \ldots, s_j)$ where $1 \leq i \leq j \leq n$.
With this context, the commit message describes the algorithm well, but the resulting algorithm can no longer be seen as a mere extension of the patience algorithm.
Also, the last sentence of the description, mentioning that the algorithm behaves like patience diff, is simply not correct.
\Cref{fig:histogram_patience} shows an example of the two algorithms giving very different results in a file that clearly contains unique lines.
In a way, the histogram algorithm could intuitively be considered a greedy variant of the patience algorithm.
\begin{figure}
    \centering
    \begin{subfigure}[b]{0.49\textwidth}
        \begin{adjustbox}{width=\textwidth,clip}
\begin{lstlisting}[language=diff]
 int cmd__reftable(int argc, const char **argv)
 {
-       /* test from simple to complex. */
        basics_test_main(argc, argv);
-       record_test_main(argc, argv);
        block_test_main(argc, argv);
-       tree_test_main(argc, argv);
-       pq_test_main(argc, argv);
-       readwrite_test_main(argc, argv);
        merged_test_main(argc, argv);
-       stack_test_main(argc, argv);
+       pq_test_main(argc, argv);
+       record_test_main(argc, argv);
        refname_test_main(argc, argv);
+       readwrite_test_main(argc, argv);
+       stack_test_main(argc, argv);
+       tree_test_main(argc, argv);
        return 0;
 }
\end{lstlisting}
        \end{adjustbox}
        \caption{Diff using the histogram algorithm}
        \label{fig:histogram_patience_a}
    \end{subfigure}
    \hfill
    \begin{subfigure}[b]{0.49\textwidth}
        \begin{adjustbox}{width=\textwidth,clip}
\begin{lstlisting}[language=diff]
 int cmd__reftable(int argc, const char **argv)
 {
-       /* test from simple to complex. */
        basics_test_main(argc, argv);
-       record_test_main(argc, argv);
        block_test_main(argc, argv);
-       tree_test_main(argc, argv);
+       merged_test_main(argc, argv);
        pq_test_main(argc, argv);
+       record_test_main(argc, argv);
+       refname_test_main(argc, argv);
        readwrite_test_main(argc, argv);
-       merged_test_main(argc, argv);
        stack_test_main(argc, argv);
-       refname_test_main(argc, argv);
+       tree_test_main(argc, argv);
        return 0;
 }
 
\end{lstlisting}
        \end{adjustbox}
        \caption{Diff using the patience algorithm}
        \label{fig:histogram_patience_b}
    \end{subfigure}
    \caption{Example that shows that histogram is not a mere extension of the patience algorithm (example from Git's source code \commitcite{fb22207}).
        All lines in the function body are unique in the file, so according to the description of the histogram algorithm, the output should be equal.
        It is easy to see that in this case, patience finds the shorter diff by computing the longest common subsequence, while the histogram algorithm greedily selects regions to pivot on, causing a longer diff.}
    \label{fig:histogram_patience}
\end{figure}
\paragraph{Behavior of the algorithm}
Another notable behavior of the algorithm is that it is neither guaranteed to select the largest common region, nor the region containing the least frequent line.
By keeping track of the current best region and updating it if \textit{either} the size is larger \textit{or} the lowest record count is lower, no guarantee can be made about the selected region.
An adversarial example could be constructed where the algorithm selects a region that has both a small size and a high lowest record count, making the algorithm pivot on an unfortunate line and causing a suboptimal diff.

Despite the apparent drawbacks, the diffs that the algorithm produces are subjectively not worse than using the other algorithms.
When reviewing diffs from Git's repository itself (included in the empirical analysis in \Cref{section:comparison}), the diffs produced are often equal to, or easier to read than, those produced by the Myers algorithm.
In most diffs observed, the histogram algorithm produces minimal length diffs.
Although there are cases where the histogram algorithm produces results very different from those produced by the patience algorithm, in most observed cases they produce very similar results.
\Cref{section:comparison} contains a more detailed comparison of the algorithms, including examples of how the diffs look more readable exactly.

\section{Pre- and Post-processing}
\label{section:prepostprocessing}
\paragraph{Preprocessing}
Before running the \texttt{minimal} or \texttt{myers} diff algorithm, Git tries to do some basic simplification of the problem.
This includes stripping common prefixes and suffixes \sourcecite{xprepare.c}{429}.
Sequences of identical lines at the beginning or end of the file will never be marked as changed and do not have to be considered by the algorithm.
It also marks lines as changed if they only occur in one of the files but not in the other (\textit{unmatched}) \sourcecite{xprepare.c}{366}.
While these two optimizations never cause non-minimal diffs, non-optimal heuristics are also applied.
In a file $A$ of length $n$, call a line $a_f$ \textit{frequent}, if and only if it occurs more than $\mathop{approxSqrt}(n)$ times in $A$.
If it is part of a contiguous block $a_x,a_{x+1}\ldots, a_f, \ldots, a_y$ where $a_x,\ldots, a_y$ are all either \textit{unmatched} or \textit{frequent} and fewer than $\frac{y-x}{4}$ of those are just \textit{frequent}, it is marked as changed \sourcecite{xprepare.c}{303}.
Intuitively, this marks lines between blocks of \textit{unmatched} lines also as changed, merging the two chunks of changed lines.
This can result in suboptimal diffs, as seen in \Cref{fig:non-minimal-diff}.
\begin{figure}
    \begin{adjustbox}{width=\textwidth,clip}
\begin{lstlisting}[language=diff]
diff --git a/4559035 b/80047b1
index 4559035c47..80047b1bb7 100644
--- a/4559035
+++ b/80047b1
@@ -340,14 +340,8 @@ extern void strbuf_vaddf(struct strbuf *sb, const char *fmt, va_list ap);
 
 /**
  * Add the time specified by `tm`, as formatted by `strftime`.
- * `tz_name` is used to expand %Z internally unless it's NULL.
- * `tz_offset` is in decimal hhmm format, e.g. -600 means six hours west
- * of Greenwich, and it's used to expand %z internally. However, tokens
- * with modifiers (e.g. %Ez) are passed to `strftime`.
- */
-extern void strbuf_addftime(struct strbuf *sb, const char *fmt,
-                           const struct tm *tm, int tz_offset,
-                           const char *tz_name);
+ */
+extern void strbuf_addftime(struct strbuf *sb, const char *fmt, const struct tm *tm);
 
 /**
  * Read a given size of data from a FILE* pointer to the buffer.
\end{lstlisting}
    \end{adjustbox}
    \caption{Non-minimal diff resulting from the \texttt{myers} option in Git.
        Note that the line \texttt{ */} is unnecessarily marked as added and removed.
        The line itself often occurs within the file and the surrounding lines of the first version are not part of the second version. The example is from Git's own repository.}
    \label{fig:non-minimal-diff}
\end{figure}
This heuristic is used with the \texttt{myers} and the \texttt{minimal} options.
While in case of the former it can be argued that it creates better-looking diffs by merging blocks of changes, with the \texttt{minimal} option selected it should be considered a bug.
This affects about 1.3\% of all diffs in the history of Git itself and was already present in the initial import of libXDiff to Git.
As part of this project I have developed a patch that disables this behavior when using the minimal option, so that the \texttt{minimal} option does indeed produce minimal diffs.
I have submitted the patch to the Git project, and it has been accepted and released as part of Git 2.50 \cite{hamanoGitV250Release2025}.

\paragraph{Post-processing}
Git does some post-processing after computing a diff to get more readable and better-looking diffs. % TODO: does it run this for all algorithms? Cite this
A notable technique is the \textit{indent heuristic}.
This is based on the fact that groups (consecutive additions or consecutive deletions) can often be \textit{slid} up or down, if the line immediately above and below the chunk are the same.
\Cref{fig:indent-heuristic} shows an example of this: while the diff does not get shorter, sliding the group one line up improves the readability substantially.
The indent heuristic was added to Git in 2009 by Michael Haggerty \cite{haggertyMhaggerDiffslidertools2025} and is still used today.
\begin{figure}
    \centering
\begin{lstlisting}[language=diff, basicstyle=\footnotesize\ttfamily]
 }
 if (!$smtp_server) {
+       $smtp_server = $repo->config('sendemail.smtpserver');
+}
+if (!$smtp_server) {
        foreach (qw( /usr/sbin/sendmail /usr/lib/sendmail )) {
 

 }
+if (!$smtp_server) {
+       $smtp_server = $repo->config('sendemail.smtpserver');
+}
 if (!$smtp_server) {
        foreach (qw( /usr/sbin/sendmail /usr/lib/sendmail )) {
\end{lstlisting}
    \caption{Example of a shifted group using the indent heuristic.
        The example is from the commit that introduced the heuristic \commitcite{433860f}.
        While both hunks have added three lines, the second one more clearly shows what was changed.}
    \label{fig:indent-heuristic}
\end{figure}

To decide on which position of a slidable group of changes is the best, the heuristic considers what is called a split, which is the position of the first and last changed line in a group.
For each split, a penalty is computed based on several factors.
Before computing the penalty, the effective indent level of a line is computed as either the amount of indent in the respective line or the first non-empty line following it.
Also, the amount of blank lines before and after the split is computed as well as a flag whether there are any blank lines surrounding the split.
Finally, several features are computed from this and a weighted average using the weights shown in \Cref{tab:indent-heuristic-weights} is computed as the penalty.
\begin{table}
    \centering
    \begin{tabular}{p{0.8\textwidth}r}
        \toprule
        \textbf{Feature}                                                              & \textbf{Weight} \\
        \midrule
        split at start of file                                                        & 1               \\
        split at end of file                                                          & 21              \\
        total number of blank lines around the split                                  & -30             \\
        number of blank lines after the split                                         & 6               \\
        line indented more than predecessor (no blanks)                               & -4              \\
        line indented more than predecessor (with blanks)                             & 10              \\
        line indented less than both predecessor and successor (no blanks)            & 24              \\
        line indented less than both predecessor and successor (with blanks)          & 17              \\
        line indented less than predecessor but not less than successor (no blanks)   & 23              \\
        line indented less than predecessor but not less than successor (with blanks) & 17              \\
        \bottomrule
    \end{tabular}
    \caption{Weights used by the indent heuristic to determine optimal position of diff hunks. \sourcecite{xdiffi.c}{547}}
    \label{tab:indent-heuristic-weights}
\end{table}
For all possible shifts of the group of changes, the penalties for the two splits are added and the shift with the lowest overall penalty is taken.
During this comparison, a factor of 60 is added to the shift that has a greater total indent (sum of the effective indents of the two splits), thus preferring shifts which split at less indented lines.
When developing the heuristic, Michael Haggerty chose these weights by manually selecting the optimal sliding position in about 6000 diffs from open source repositories and then optimizing the parameters to match the manually selected diffs as closely as possible \cite{haggertyMhaggerDiffslidertools2025}.
According to the analysis done at the time, Git previously chose the human-rated version in about 40\% of diffs, while with the heuristic it does so in over 97\% of diffs \commitcite{433860f}.

\section{Empirical Comparison}
\label{section:comparison}
To compare the algorithms, multiple different metrics can be used.
As stated before, the length of the generated diff (number of lines added or removed) by a diffing algorithm is one of the most fundamental metrics for a diffing algorithm.
Still, longer diffs are not necessarily worse and can be much more readable than shorter diffs (as seen in \Cref{fig:histogram_good}).
The minimal and Myers algorithms are fundamentally based on minimizing the number of changes, while the histogram and patience algorithms do not and prioritize readability.
Another important metric is the performance of the diffing algorithms, which seems to be a major reason for the introduction of the histogram algorithm \commitcite{8c912eea}.

\paragraph{Experimental Setup}
To compare the different algorithms empirically, I ran them on all commits in the repository of Git.
I excluded any merge commits and compared the tree of a commit with the tree of its parent commit using libGit2 to identify the blobs that have been changed.
At the time of the evaluation, this resulted in 163,974 distinct diffs (pairs of two blobs), which were part of 62,139 commits.
This difference arises from the fact that commits can affect multiple files and with tools like rebases and cherry-picks, a diff can be part of multiple commits.
Since I only compare the algorithms for computing differences between two files, I did not record changes in permissions and filenames.
However, libGit2 does use a rename detection heuristic when comparing two trees to identify the blobs that have been changed.
I ran these experiments on a modified version of Git 2.49, in which I had fixed the aforementioned bug in the \textit{minimal} option and added instrumentation to measure the execution time of the diff algorithms with high precision.
This way, the execution time of the diffing algorithm can be measured significantly more accurately, without the overhead and variance introduced by the process creation.
To further reduce the variance, I have run the diffs single-threaded and averaged over 15 runs.
For reference, I ran the experiments on an AMD Ryzen 7 7500U CPU with 16 GB of RAM, using Ubuntu 22.04.
The code for the experiments is available on GitHub (\url{https://github.com/yndolg/GitDiffAndMerge}).

\begin{figure}
    \centering
    \includegraphics[width=\textwidth]{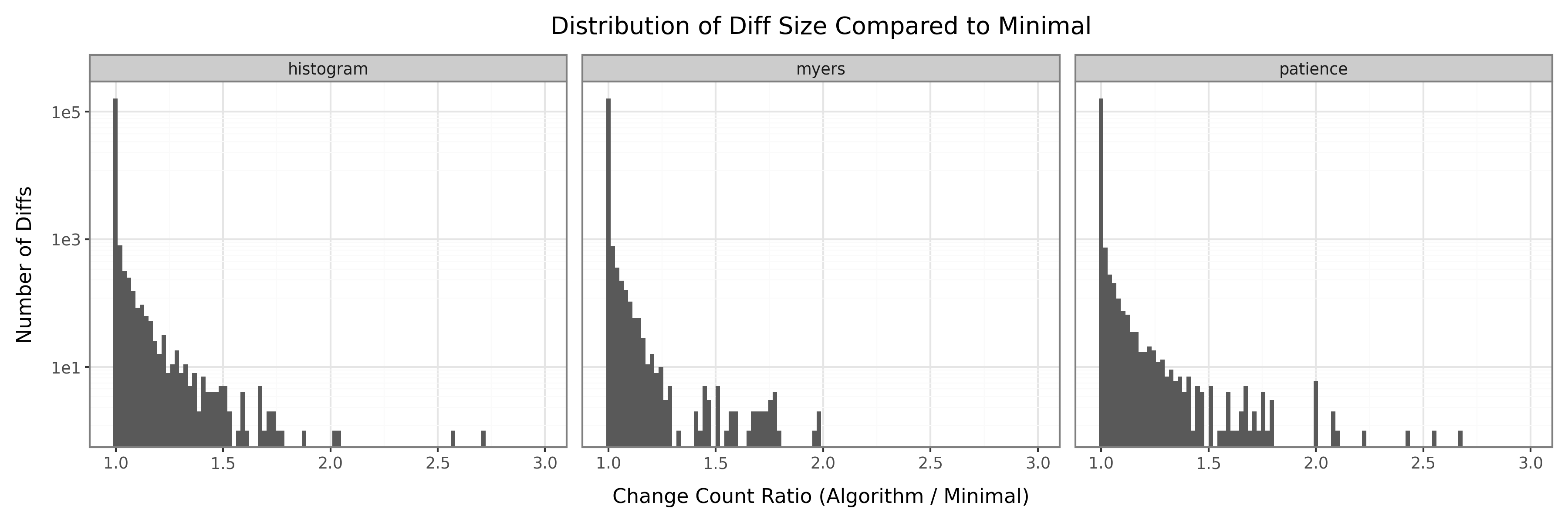}
    \caption{Histogram of the size increase of the patience, Myers and histogram algorithms over the minimal algorithm. Note that the y-axis is logarithmic.}
    \label{fig:diff_comparison_size}
\end{figure}

\Cref{fig:diff_comparison_size} and \Cref{tab:diff_comparison_size} show the size of the diffs generated by the myers, patience, and histogram algorithms relative to the diffs generated by the minimal algorithm.
It can be seen that for most diffs, the algorithms all produce a minimal diff and only a small fraction of diffs is larger using the other algorithms.
All three non-minimal algorithms have a similar size distribution.
\begin{table}
    \centering
    \begin{tabular}{rll}
        \toprule
        Algorithm & Frequency of larger diffs & Average increase in size (for larger diffs) \\ \midrule
        Histogram & 1.73\% $\pm$ 0.06\%       & 5.48 \% $\pm$ 0.3 \%                        \\
        Myers     & 1.54\% $\pm$ 0.06\%       & 4.73 \% $\pm$ 0.3 \%                        \\
        Patience  & 1.56\% $\pm$ 0.06\%       & 5.71 \% $\pm$ 0.4 \%                        \\
        \bottomrule
    \end{tabular}
    \caption{Size increase over the minimal diff for the different algorithms.
        The first column shows for how many diffs the algorithm produces larger-than-minimal results.
        The second column shows the geometric mean of the percentage increase in size (only for diffs that are larger than minimal).
        The given errors are the size of the 95\% confidence interval. }
    \label{tab:diff_comparison_size}
\end{table}
They produce non-minimal diffs in only about 1.6\% of diffs, with the Myers and patience algorithms being optimal slightly---but statistically significantly---more often than the histogram algorithm.
In the cases where the diffs are non-minimal, all three algorithms produce about 5\% larger diffs on average.

\begin{figure}
    \centering
    \includegraphics[width=\textwidth]{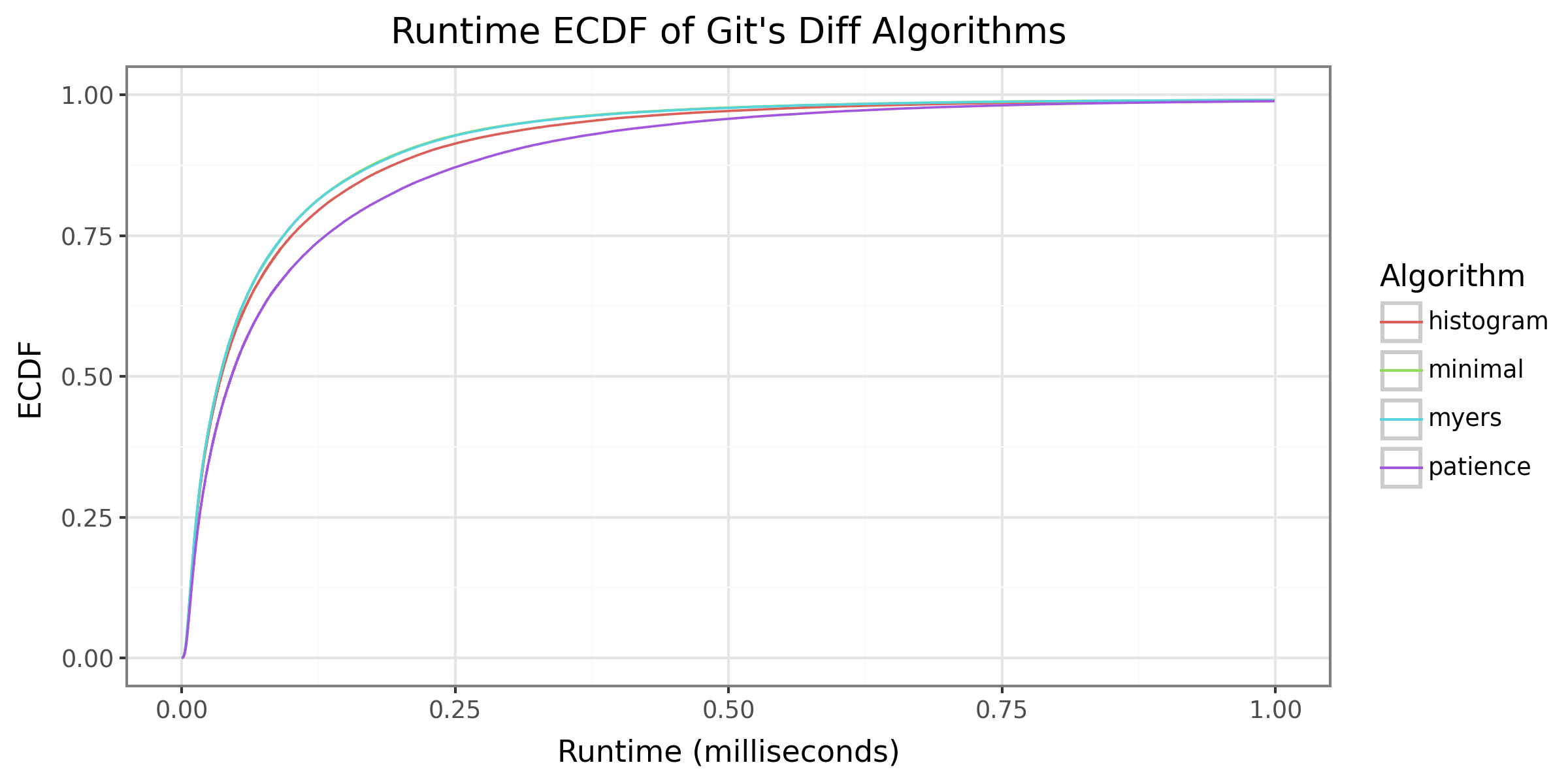}
    \caption{Distribution of the execution times of the diff algorithms as implemented in Git.}
    \label{fig:diff_comparison_timing}
\end{figure}

The runtime distributions of the different algorithms are shown in \Cref{fig:diff_comparison_timing}.
\Cref{tab:diff_comparison_timing} summarizes the timing distributions.
These timings refer to the time to compute a single diff between two files, not to compute the diff of an entire commit.
The timings also exclude the time required to start the process, read the files from disk and output the diff.
\begin{table}
    \centering
    \begin{tabular}{rll}
        \toprule
        Algorithm & Average Runtime (ms) & 95th Percentile (ms) \\ \midrule
        Minimal   & 0.117 $\pm$ 0.0021   & 0.31                 \\
        Myers     & 0.101 $\pm$ 0.0005   & 0.31                 \\
        Histogram & 0.115 $\pm$ 0.0005   & 0.36                 \\
        Patience  & 0.128 $\pm$ 0.0004   & 0.46                 \\
        \bottomrule
    \end{tabular}
    \caption{Average and 95th percentile runtime of the different diff algorithms in Git. For the average, the 95\% confidence interval is given.}
    \label{tab:diff_comparison_timing}
\end{table}
Overall runtimes for a single diff on a modern computer are very low, with the 95th percentile for all algorithms being below half a millisecond.
The distribution shows that, while all algorithms have a similar runtime, the patience algorithm is less often very fast with a runtime below 0.1 ms.
This is also reflected in a higher 95th percentile runtime than the other algorithms.
On average, the Myers algorithm is the fastest, followed by the histogram and minimal algorithms with the patience algorithm being the slowest.
While the differences are also statistically significant, overall they are very small and should not be a major factor when deciding which algorithm to use.
With runtimes this close and overall low, the choice of the diff algorithm should primarily be based on the quality of the resulting diffs.

\section{Interesting Examples}
To find more subjective differences in the quality of the diffs, I have manually examined some diffs of the empirical data set.
In cases where few changes were made, with large blocks of unchanged lines in between, most of the time all four algorithms produce the same diff.
I have found more interesting cases by specifically looking at the diffs where the number of lines changed differs significantly between the algorithms.
These often include either blank lines or a reordering of lines (either by moving an entire block or by re-ordering individual lines).
These are also the cases where many distinct diffs are possible, and it might be hard to decide which one is the best.
The diff shown in \Cref{fig:histogram_patience} demonstrates this problem well, as it is a pure reordering of lines.
In that example, the minimal and Myers diffs both result in the same diff as the patience algorithm does, since they are also based on the longest common subsequence.

\begin{figure}
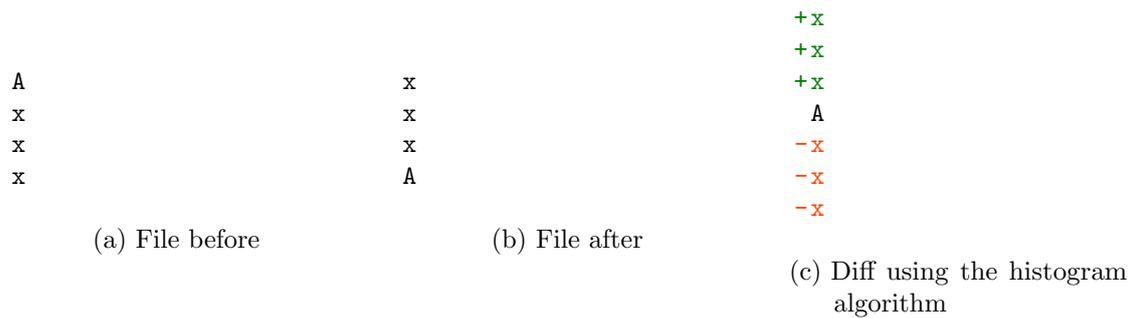

    \begin{subfigure}[c]{0.3\textwidth}
\begin{lstlisting}[language=diff]
A
x
x
x
\end{lstlisting}
        \caption{File before}
    \end{subfigure}
    \hfill
    \begin{subfigure}[c]{0.3\textwidth}
\begin{lstlisting}[language=diff]
x
x
x
A
\end{lstlisting}
        \caption{File after}
    \end{subfigure}
    \hfill
    \begin{subfigure}[c]{0.3\textwidth}
\begin{lstlisting}[language=diff]
+x
+x
+x
 A
-x
-x
-x
\end{lstlisting}
        \caption{Diff using the histogram algorithm}
    \end{subfigure}
    \caption{Example of the greediness of the histogram algorithm.}

    \label{fig:histogram_bad}
\end{figure}

In contrast, the histogram algorithm greedily selects one of the matching lines and pivots on it, potentially resulting in a large diff.
This behavior can be exploited to generate highly suboptimal diffs, like the one in \Cref{fig:histogram_bad}.
In that case, the algorithm selects the \texttt{A} line as the pivot, which was moved from the top to the bottom of the file, requiring a large diff (the example works with arbitrarily many other lines in between).
I have seen this behavior in the real-world commits, particularly when a block of code is moved around.
Note that this example also demonstrates that the histogram algorithm is not symmetric.
Running the reverse of the diff results in the more intuitive diff with just two lines changed.
This is because the histogram algorithm is not symmetric in its selection of the pivot (\texttt{scanA} only scans the old file, not the new one).
Both Myers and minimal produce the smaller diff, while patience produces the same one as the histogram algorithm (since A is the only unique line in the file).
However, similar examples can be constructed where patience uses shorter diffs than histogram, like in \Cref{fig:histogram_patience}.

However, when subjectively examining the diffs, I have found that the histogram algorithm produces very readable diffs in many cases.
An example of this is shown in \Cref{fig:histogram_good}.
In that case, while not adding and deleting the closing brace enlarges the diff, it is much easier to understand as the whole block of code is moved.
\begin{figure}
    \begin{subfigure}[c]{\textwidth}
\begin{lstlisting}[language=diff, basicstyle=\tiny\ttfamily]
        if (verbose || /* Truncate the message just before the diff, if any. */
            cleanup_mode == CLEANUP_SCISSORS)
                strbuf_setlen(&sb, wt_status_locate_end(sb.buf, sb.len));
+
        if (cleanup_mode != CLEANUP_NONE)
                strbuf_stripspace(&sb, cleanup_mode == CLEANUP_ALL);
-
-       if (message_is_empty(&sb) && !allow_empty_message) {
-               rollback_index_files();
-               fprintf(stderr, _("Aborting commit due to empty commit message.\n"));
-               exit(1);
-       }
        if (template_untouched(&sb) && !allow_empty_message) {
                rollback_index_files();
                fprintf(stderr, _("Aborting commit; you did not edit the message.\n"));
                exit(1);
        }
+       if (message_is_empty(&sb) && !allow_empty_message) {
+               rollback_index_files();
+               fprintf(stderr, _("Aborting commit due to empty commit message.\n"));
+               exit(1);
+       }
 
        if (amend) {
                const char *exclude_gpgsig[2] = { "gpgsig", NULL };
\end{lstlisting}
        \caption{Histogram and patience diff}
        \label{fig:histogram_good_histogram}
    \end{subfigure}
    \begin{subfigure}[c]{\textwidth}
\begin{lstlisting}[language=diff, basicstyle=\tiny\ttfamily]
        if (verbose || /* Truncate the message just before the diff, if any. */
            cleanup_mode == CLEANUP_SCISSORS)
                strbuf_setlen(&sb, wt_status_locate_end(sb.buf, sb.len));
+
        if (cleanup_mode != CLEANUP_NONE)
                strbuf_stripspace(&sb, cleanup_mode == CLEANUP_ALL);
-
-       if (message_is_empty(&sb) && !allow_empty_message) {
+       if (template_untouched(&sb) && !allow_empty_message) {
                rollback_index_files();
-               fprintf(stderr, _("Aborting commit due to empty commit message.\n"));
+               fprintf(stderr, _("Aborting commit; you did not edit the message.\n"));
                exit(1);
        }
-       if (template_untouched(&sb) && !allow_empty_message) {
+       if (message_is_empty(&sb) && !allow_empty_message) {
                rollback_index_files();
-               fprintf(stderr, _("Aborting commit; you did not edit the message.\n"));
+               fprintf(stderr, _("Aborting commit due to empty commit message.\n"));
                exit(1);
        }
\end{lstlisting}
        \caption{Myers and minimal diff}
        \label{fig:histogram_good_myers}
    \end{subfigure}
    \caption{Example of a longer diff that looks subjectively more readable.
        \Cref{fig:histogram_good_histogram} is the diff output by both the histogram and patience algorithms, while \Cref{fig:histogram_good_myers} is the diff output by the Myers and minimal algorithms.
        The example can be reproduced as \texttt{git diff e7a2cb62 4bbac014} in the Git repository of Git itself.}
    \label{fig:histogram_good}
\end{figure}
The pattern that can be observed is that in a minimal (or Myers) diff, often less relevant lines like empty lines or closing braces are used to create a smaller diff.
This results in the old code and new code being interwoven in the diff at irrelevant lines, while it would be much cleaner to show the entire block as added or deleted, as the patience and histogram algorithms often do.

A phenomenon that is much easier to produce using the minimal and Myers algorithms is that they are not local.
Inserting or deleting a line at the start of the file might cause the algorithm to choose a different diff for a change at the end of the file.
This might not seem like a major problem, but it does lead to problems when using the diff as part of a three-way merge.

\section{Chapter Summary}

This chapter has examined the four diff algorithms available in Git: Myers, minimal, patience, and histogram.
While the Myers and minimal algorithms are both based on computing longest common subsequences to minimize diff length, the patience and histogram algorithms focus on unique lines instead.
I have described the histogram algorithm, which is the default for merges and considered the most modern diff algorithm, and found that it behaves significantly differently than previously believed.
In fact, it is not based on longest common subsequences at all, but rather a greedy algorithm based on line frequency.

The empirical analysis reveals that runtime performance differences between the algorithms are negligible in practice, with all algorithms completing most diffs in well under a millisecond.
The choice of algorithm should therefore be based primarily on diff quality rather than performance considerations.
In terms of diff size, all non-minimal algorithms produce optimal results in approximately 98.8\% of cases, with only modest increases in size when they deviate from optimality.
Subjectively, the patience and histogram algorithms produce highly readable diffs, since they less often interweave unrelated lines.

The fact that both the patience and histogram algorithms are not solely based on minimizing the edit script length allows the construction of highly pathological examples.
I have shown an example where, when moving a single line, the entire rest of the file is marked as changed, which can also be observed in practice.
These cases primarily arise when code is moved or re-ordered.
Furthermore, the histogram algorithm is not symmetric.
Running a diff in reverse might yield a significantly larger or smaller diff.

\chapter{Merge}
\label{section:merge}
The second major operation in Git is merging, which combines two or more sets of changes that were developed independently of each other.
In this section, I will first describe how merges in Git are performed and then discuss some (unexpected) implications of this process.

\section{Merge Strategies}
\label{section:merging_strategies}
As described in \Cref{section:datastructure_merge}, a merge first has to identify a common ancestor of the two branches before it can then perform a three-way merge.
Notice that a merge is only concerned with these three versions: the common ancestor $O$, the most recent version of the left branch $L$, and the most recent version of the right branch $R$.
The exact history of the branches is not relevant for the merge itself, but only for determining the common ancestor.
The exact procedure of how to determine the common ancestor as well as determining which files require a three-way merge is defined by a \textit{merge strategy}  \cite{GitMergestrategiesDocumentation}.
The strategies available in Git are:
\begin{description}
    \item[resolve] Legacy strategy that, in case there is more than one common ancestor, chooses one as a base arbitrarily. 
        It is still available for backwards compatibility.
    \item[recursive] More advanced version of resolve that introduced recursive merging of the common ancestors.
        This strategy was the default when merging two branches up to Git 2.34 (released in November 2021) \cite{blauHighlightsGit2342021}.
        In Git 2.50 (released in June 2025)\cite{blauHighlightsGit2502025}, it was removed and is now an alias for the \texttt{ort} strategy.
    \item[ort] The \texttt{ort} strategy was developed as a direct replacement for the \texttt{recursive} strategy.
        Its overall structure is similar to \texttt{recursive}, but it improves some edge cases, is significantly faster, and improves the code quality \cite{blauHighlightsGit2332021}.
        This merge strategy is used for most merges and is discussed below in more detail.
    \item[octopus] This strategy allows for a merge of more than two branches in a single merge commit (which resolve, \texttt{recursive}, and \texttt{ort} cannot perform).
        It is supposed to be used when merging multiple feature branches without creating multiple merge commits.
        The conflict resolution of the octopus strategy is limited: I have confirmed that it performs a three-way merge if a file has been modified on multiple branches, but does not allow for manual conflict resolution.
        If a conflict occurs, the merge fails and leaves the repository untouched.
    \item[ours] Instead of running a proper merge, this strategy simply reuses the current tree, completely ignoring the changes done on the other branch.
    \item[subtree] A modified \texttt{ort} strategy used to merge a branch with a subdirectory of a repository.
        It allows pulling changes from a dependency that was previously imported into a subdirectory \cite[p. 286]{chaconProGit2014}.
\end{description}

\paragraph{The \texttt{ort} Strategy}
The \texttt{ort} strategy has been developed as a replacement for the recursive strategy.
The motivation for the rewrite was the poor code quality of the recursive strategy implementation and performance issues regarding file rename detection.
Merges with the recursive strategy containing a large number of renames could take as long as 30 minutes, which becomes an even larger problem when repeatedly merging during a rebase \cite{newrenOptimizingGitsMerge2022}.
The name \textit{ort} is an acronym (``Ostensibly Recursive's Twin'') and was chosen because together with the flag for selecting the merge strategy (\texttt{-s}), the command for running it contains \texttt{-sort} \sourcecite{merge-ort.c}{12}.

\begin{figure}
    \centering
    \begin{tikzpicture}[
    commit/.style={circle, draw=black, minimum size=6mm},
    >=Stealth
  ]
  % Nodes (commits)
  \node[commit] (C1) at (0,0) {$C_1$};
  \node[commit] (C2) at (2,0) {$C_2$};
  \node[commit] (C3) at (0,3) {$C_3$}; % Feature branch
  \node[commit] (C4) at (2,3) {$C_4$}; % Feature branch continues

  \node[commit] (m) at (4, 1.5) {$?$};

  % Edges (arrows showing commit progression)
  \draw[->] (C1) -- (C2);
  \draw[->] (C2) -- (m);
  \draw[->] (C3) -- (C4);
  \draw[->] (C4) -- (m);
  \draw[->] (C1) -- (C4);
  \draw[->] (C3) -- (C2);

  \draw[->] (-1,3) -- (C3);
  \draw[->] (-1,0) -- (C1);

  \begin{scope}[xshift=7cm]
    % Nodes (commits)
    \node[commit] (C1) at (0,0) {$C_1$};
    \node[commit] (C2) at (2,0) {$C_2$};
    \node[commit] (C3) at (0,3) {$C_3$}; % Feature branch
    \node[commit] (C4) at (2,3) {$C_4$}; % Feature branch continues

    \node[commit, draw=none, fill=black!20] (CV) at (2.6, 1.5) {$C_v$}; % Common ancestor
    \node[commit] (m) at (4, 1.5) {$C_5$};

    % Edges (arrows showing commit progression)
    \draw[->] (C1) -- (C2);
    \draw[->] (C2) -- (m);
    \draw[->] (C3) -- (C4);
    \draw[->] (C4) -- (m);
    \draw[->] (C1) -- (C4);
    \draw[->] (C3) -- (C2);

    \draw[->] (-1,3) -- (C3);
    \draw[->] (-1,0) -- (C1);

    \draw[->, draw=black!50] (C1) -- (CV);
    \draw[->, draw=black!50] (C3) -- (CV);
    \draw[dashed, draw=black!50] (CV) -- (m);
  \end{scope}
\end{tikzpicture}
    \caption{Virtual commits in the \texttt{ort} strategy. The left side shows a commit graph with a non-unique lowest common ancestor ($C_1$ and $C_3$ are lowest common ancestors).
        The \texttt{ort} strategy first merges all lowest common ancestors into a virtual commit ($C_v$), which will then be used as a base for the three-way merge.
        The virtual commit will not be part of the commit history.}
    \label{fig:recursive_merge}
\end{figure}
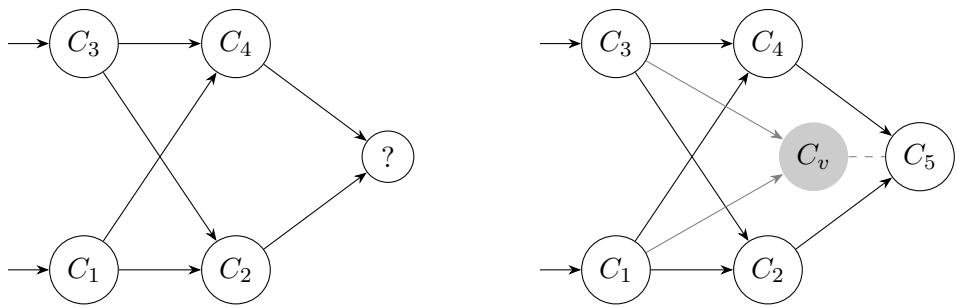

An important part of the \texttt{ort} strategy (which was previously part of the recursive strategy) is its procedure for determining the common ancestor for a three-way merge.
Since the lowest common ancestor of two nodes in a directed acyclic graph does not have to be unique, this is a more nuanced question than it might seem.
\Cref{fig:recursive_merge} shows such a situation, named a \textit{crisscross merge}, where multiple lowest common ancestors exist.
The \texttt{ort} strategy solves this problem by merging the lowest common ancestors into a virtual commit, which is then used as the base for the three-way merge.
This commit is not written to disk and will not become part of the commit graph.
When merging the common ancestors, the same problem can occur again, so the \texttt{ort} strategy will be applied recursively.
\Cref{alg:ort_merge} gives pseudocode for this procedure.
\begin{algorithm}
\caption{Pseudocode for a merge of two commits $C_1$ and $C_2$ using the \texttt{ort} strategy.}
\label{alg:ort_merge}
\SetAlgoLined
\SetKwFunction{Merge}{merge}
\SetKwFunction{FindLCA}{findLowestCommonAncestors}
\SetKwFunction{Reverse}{reverse}
\SetKwFunction{ThreeWayMerge}{threeWayMerge}
\SetKwData{Ancestors}{ancestors}
\SetKwProg{Fn}{Function}{:}{}
\SetKwData{Merged}{merged}
\Fn{\Merge{$C_1$,$C_2$}}{
 \Ancestors $\leftarrow$ \FindLCA{$C_1$, $C_2$} \\
 \tcc{\Ancestors is ordered by creation time, oldest first}
 \Merged $\leftarrow$ \Ancestors.\textit{getAndRemoveLast}() \\
 \While {$\left|\Ancestors\right| > 0$}{ 
    \Merged $\leftarrow$ \Merge{\Merged, \Ancestors.\textit{getAndRemoveLast}()} \\
}
    \Return \ThreeWayMerge{$C_1$,\Merged ,$C_2$} \\
}
\vspace{0.5cm}
\end{algorithm}
The ancestors of a merge are merged into the virtual commit in sequence, ordered by descending creation time.
This has been reported to result in fewer conflicts than not ordering the ancestors before merging \sourcecite{merge-ort.h}{l. 105}.
Since the merges into the virtual commit might themselves have multiple lowest common ancestors, the procedure is repeated recursively.

\section{Exponential Merge}
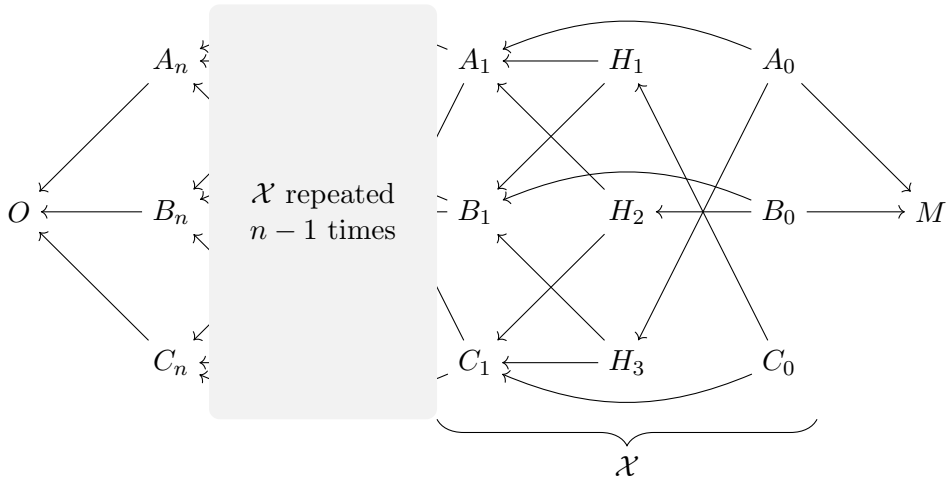
\begin{figure}
    \centering
    \begin{tikzpicture}
    \node (a1) at (0, -0) {$A_0$};
    \node (b1) at (0, -2) {$B_0$};
    \node (c1) at (0, -4) {$C_0$};

    \node (h1) at (-2, -0) {$H_1$};
    \node (h2) at (-2, -2) {$H_2$};
    \node (h3) at (-2, -4) {$H_3$};

    \node (a2) at (-4, -0) {$A_1$};
    \node (b2) at (-4, -2) {$B_1$};
    \node (c2) at (-4, -4) {$C_1$};

    \node (h3_2) at (-6, -0) {$H_3$};
    \node (h4) at (-6, -2) {$H_4$};
    \node (h4_2) at (-6, -4) {$H_4$};

    \node (a3) at (-8, -0) {$A_n$};
    \node (b3) at (-8, -2) {$B_n$};
    \node (c3) at (-8, -4) {$C_n$};

    \draw[->] (a1) -- (h3);
    \draw[->] (b1) -- (h2);
    \draw[->] (c1) -- (h1);

    \draw[->] (h1) -- (a2);
    \draw[->] (h3) -- (c2);
    \draw[->] (h1) -- (b2);
    \draw[->] (h2) -- (a2);
    \draw[->] (h2) -- (c2);
    \draw[->] (h3) -- (b2);

    \draw[->] (a2) -- (h4_2);
    \draw[->] (b2) -- (h4);
    \draw[->] (c2) -- (h3_2);

    \draw[->] (h3_2) -- (a3);
    \draw[->] (h4_2) -- (c3);
    \draw[->] (h3_2) -- (b3);
    \draw[->] (h4) -- (a3);
    \draw[->] (h4) -- (c3);
    \draw[->] (h4_2) -- (b3);

    \draw[->] (a1) to[bend right=23] (a2);
    \draw[->] (b1) to[bend right=23] (b2);
    \draw[->] (c1) to[bend left=23] (c2);

    \draw[->] (a2) to[bend right=23] (a3);
    \draw[->] (b2) to[bend right=23] (b3);
    \draw[->] (c2) to[bend left=23] (c3);

    \draw[decorate,decoration={brace,amplitude=10pt}] 
        (0.5,-4.75) --node[midway,below=10pt] {$\mathcal{X}$} (-4.5,-4.75);
        
    \draw[rectangle, draw=none, fill=gray!10, rounded corners] 
        (-7.5,-4.75) rectangle (-4.5,0.75);
    \node[align=center] at (-6,-2) {$\mathcal{X}$ repeated \\ $n-1$ times};

    \node (o) at (-10, -2) {$O$};
    \draw[->] (a3) -- (o);
    \draw[->] (b3) -- (o);
    \draw[->] (c3) -- (o);

    \node (m) at (2, -2) {$M$};
    \draw[->] (a1) -- (m);
    \draw[->] (b1) -- (m);

\end{tikzpicture}
    \caption{An example of a merge requiring exponential time in Git. 
    Merging $A_0$ and $B_0$ results in a large number of recursive calls, with each additional block $\mathcal{X}$ doubling the number of recursive calls required.}
    \label{fig:exponential_merge}
\end{figure}

Since the procedure for creating the merge base of the \texttt{ort} strategy is recursive (\Cref{alg:ort_merge}), it has the potential to exhibit exponential time complexity.
In fact, I have been able to construct a commit graph of $V = 6n + 4$ commits, in which a merge of two branches can be run that results in $T(n) = 2^{n}+1$ executions of the \texttt{merge} function in \Cref{alg:ort_merge}.
This example can be seen in \Cref{fig:exponential_merge}.
The block $\mathcal{X}$ is constructed in such a way that the lowest common ancestors of $A_i$ and $B_i$ are $\operatorname{lca}(A_i, B_i) = \left\{A_{i+1}, B_{i+1}, C_{i+1}\right\}$, which have to be merged together.
After the merge of $A_{i+1}$ and $B_{i+1}$ has been created, it has to be merged with $C_{i+1}$.
For this merge, the lowest common ancestors are again the set $\left\{A_{i+2}, B_{i+2}, C_{i+2}\right\}$, which get merged together another time.
Merging $A_i$, $B_i$, and $C_i$ results in $A_{i+1}$, $B_{i+1}$, and $C_{i+1}$ being merged together into a virtual commit twice, which in turn requires merging $A_{i+1}$, $B_{i+2}$, and $C_{i+2}$ into a virtual commit four times.
This series continues and requires exponentially many merges in $n$.

I have confirmed this behavior with a script that creates the commit graph as in \Cref{fig:exponential_merge} and runs a merge between $A_0$ and $B_0$.
The repository only contained a single file and all commits except the initial commit were empty.
With $n = 17$, the merge between $A_0$ and $B_0$ already took 66 seconds, roughly doubling with each additional block.

\section{Three-Way Merge}
\label{section:three_way_merge}
When the merge strategy has determined that a file was changed in two branches ($L$ and $R$), and has determined a common ancestor $O$ to use, Git needs to analyze the content of the files and perform the merge.
This procedure is called the \textit{three-way merge} (also known as \textit{diff3}, after the UNIX utility \cite{GNUDiffutils}).
It first runs a diff algorithm to find the differences between $O$ and $L$, as well as the differences between $O$ and $R$.
Then, the three-way merge algorithm is used to apply both changes to the ancestor $O$.
This will result in a version that has both the changes from $L$ and the changes from $R$ -- a merge of $L$ and $R$.

The core three-way merge procedure is only concerned with applying the changes from $O$ to $L$ and $O$ to $R$ to $O$.
In Git, this is only possible if the files are text files.
If a binary file is modified in both branches, this will always result in a conflict.

\paragraph{A Historical Note}
In the beginning, Git relied on external tools for running three-way merges.
Namely, it used the \texttt{merge} command available as part of Revision Control System (GNU RCS) \cite{tichyRCSSystemVersion1985} \commitsourcecite{278fcd7}{git-merge-one-file.sh}{107}.
Revision Control System in turn relied on the \texttt{diff3} command that is part of GNU diffutils to run the three-way merge.
The \texttt{diff3} command in GNU diffutils was originally developed by Randy Smith (the commit is dated to 1992) and contains the core algorithm.
The output format was also already present in GNU diffutils.
Today, Git has its own implementation of a three-way merge in its fork of the libXDiff library.
While it was implemented from scratch, it draws heavy inspiration from the \texttt{diff3} command of libXDiff \commitcite{857b933}.
Since the initial implementation in Git, the algorithm has seen only minor changes and is still essentially the same.

\subsection{Core Algorithm}
\label{section:merge_core_algorithm}
I have derived the following description of the three-way merge algorithm from the source code of Git \sourcecite{xmerge.c}{}.
The algorithm starts by computing diffs between the ancestor file $O$ and both changed versions $L$ and $R$.
This is done using the same algorithms (by default histogram) and options that are available when running single diffs \sourcecite{xmerge.c}{695}.
The diffs are converted to the edit script format (from the changed lines format) and the entire algorithm operates on the resulting hunks.

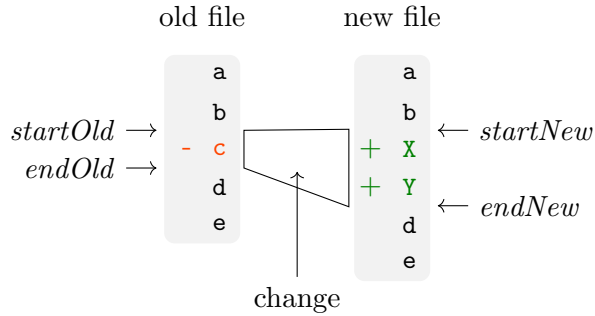
\begin{figure}
    \centering
    \begin{tikzpicture}[
        source/.style={font=\small\ttfamily, anchor=west },
    ]
    \draw[fill=gray!10, rounded corners, draw=none] (-0.5,0.25) rectangle (0.5, -2.25);
    \node[above] at (0, 0.5) {old file};

    \node[source] (a1) at (0, 0) {a};
    \node[source] (a2) at (0, -0.5) {b};
    \node[source, text=diffrem] (a3) at (0, -1) {c};
    \node[anchor=east, text=diffrem] (a3b) at (a3.west) {-};

    \draw[fill=gray!10, rounded corners, draw=none] (2, 0.25) rectangle (3, -2.75);
    \node[above] at (2.5, 0.5) {new file};

    \node[source] (a4) at (0, -1.5) {d};
    \node[source] (a5) at (0, -2) {e};

    \node[source] (b1) at (2.5, 0) {a};
    \node[source] (b2) at (2.5, -0.5) {b};
    \node[source, text=diffincl] (b3) at (2.5, -1) {X};
    \node[anchor=east, text=diffincl] (b3b) at (b3.west) {+};
    \node[source, text=diffincl] (b4) at (2.5, -1.5) {Y};
    \node[anchor=east, text=diffincl] (b4b) at (b4.west) {+};

    \node[source] (b5) at (2.5, -2) {d};
    \node[source] (b6) at (2.5, -2.5) {e};

    \draw[draw=black] (0.55, -0.75) -- (b3b.north west) -- (b4b.south west) -- (0.55, -1.25) -- cycle;

    \node (hunkLabel) at (1.25, -3) {change};
    \draw[->] (hunkLabel.north) -- (1.25, -1.3);

    \node[anchor=east] (startOld) at (-1, -0.75) {\textit{startOld}};
    \draw[->] (startOld.east) -- ++(0.4, 0);

    \node[anchor=west] (startNew) at (3.5, -0.75) {\textit{startNew}};
    \draw[->] (startNew.west) -- ++(-0.4, 0);

    \node[anchor=east] (endOld) at (-1, -1.25) {\textit{endOld}};
    \draw[->] (endOld.east) -- ++(0.4, 0);

    \node[anchor=west] (endNew) at (3.5, -1.75) {\textit{endNew}};
    \draw[->] (endNew.west) -- ++(-0.4, 0);
\end{tikzpicture}
    \caption{Visualization of a change (hunk). 
    In this example, the line \texttt{c} has been removed and the lines \texttt{X} and \texttt{Y} have been added in the same position.
    This change can be represented by the four line numbers \textit{startOld}, \textit{endOld}, \textit{startNew}, and \textit{endNew}.}
    \label{fig:hunk_example}
\end{figure}

A \textit{change} $C$ (or hunk, but called change to keep consistency with the source code) means that the lines $C.\mathit{startOld}$ to $C.\mathit{endOld}$ in the old version have been changed to the lines $C.\mathit{startNew}$ to $C.\mathit{endNew}$ in the new version.
The old version will always be the ancestor file $O$, while the new version will be either $L$ or $R$.
Intuitively, this represents a deletion, addition, or modification of multiple lines.
An example of this format is shown in \Cref{fig:hunk_example}.

\paragraph{Merge Regions}
The merge algorithm operates on these changes and generates \textit{merge regions}.
A merge region $M$ represents that the lines $M.\mathit{startAncestor}$ to $M.\mathit{endAncestor}$ in the ancestor have been changed to the lines $M.\mathit{startLeft}$ to $M.\mathit{endLeft}$ in $L$ and to the lines $M.\mathit{startRight}$ to $M.\mathit{endRight}$ in $R$.
This can be seen as a change, but for all three files.
From these merge regions, the final output can be generated.
If a merge region contains exactly the same text in both $L$ and $O$, but different text in $R$, the merge recognizes this as a change in $R$ and outputs only the changed lines from $R$ instead of the lines in $O$ (the same applies symmetrically for $L$).
If the merge region is equal in $L$ and $R$, but different in $O$, this means that the same change was applied in both branches.
This case is called a \textit{false conflict} and is not output as a conflict, but instead just applied.
The final case occurs when the content in $L$, $O$, and $R$ is all different.
In this case, a conflict occurred and the merge region is output as a conflict.
The areas between the merge regions were not changed in any file and are always the same in all three files.
Git takes them from the left file when generating the merged file.
The core procedure of the algorithm is concerned with finding these merge regions from the hunks (changes) of the two diffs.

The core loop always analyzes the first remaining change of each file ($C_L$ and $C_R$) and does a case distinction \sourcecite{xmerge.c}{505} (I have visualized these cases in \Cref{fig:merge}):
\begin{enumerate}
    \item $C_L.\mathit{endOld} < C_R.\mathit{startOld}$: \label{item:merge_case_1} In the ancestor, the left change $C_L$ touches only lines that come before any lines that are changed in the right change $C_R$.
          Thus, the change $C_L$ can be applied without conflicts.
          A merge region $M$ is created with
          \begin{align*}
            M.\mathit{startAncestor} &= C_L.\mathit{startOld} \\
            M.\mathit{endAncestor} &= C_L.\mathit{endOld} \\
            M.\mathit{startLeft} &= C_L.\mathit{startNew} \\
            M.\mathit{endLeft} &= C_L.\mathit{endNew} 
          \end{align*}
          To determine the position of this section in the right file, the algorithm looks backwards from the next change in the right file:
          \begin{align*}
            M.\mathit{startRight} &= C_R.\mathit{startNew}- C_R.\mathit{startOld} + C_L.\mathit{startOld} \\
            M.\mathit{endRight} &= C_R.\mathit{endNew} - C_R.\mathit{startOld} + C_L.\mathit{startOld} 
          \end{align*}
          As there are no changes before $C_R$, the section of $M$ in the ancestor and right file will be identical.
          After the new merge region has been recorded, the change $C_L$ is removed from the list of changes.
    \item $C_R.\mathit{endOld} < C_L.\mathit{startOld}$: \label{item:merge_case_2} This is the symmetric case to \ref{item:merge_case_1}.
          The change $C_R$ can be applied without conflicts.
          The merge region is computed analogously to the previous case.
    \item \label{item:merge_case_conflict}If the previous two cases did not apply, and 
        \begin{align*}
            & L[C_L.\mathit{startNew}..C_L.\mathit{endNew}] \neq R[C_R.\mathit{startNew}..C_R.\mathit{endNew}]\\ 
            &\lor C_L.\mathit{startOld} \neq C_R.\mathit{startOld} \\
            &\lor C_L.\mathit{endOld} \neq C_R.\mathit{endOld},
        \end{align*}
          a conflict has been detected.
          The conflicting region has to be expanded to include, in both the left and right files, all lines which are part of the two changes, not just the change on the respective side.
          To do this, the algorithm computes the start and end offset in the ancestor file: $o_s = C_L.\mathit{startOld} - C_R.\mathit{startOld}$ and $o_e = C_L.\mathit{endOld} - C_R.\mathit{startOld}$.
          For the rest of the computation, assume without loss of generality that $o_s \geq 0$.
          The merge region $M$ can be computed as 
          \begin{align*}
            M.\mathit{startAncestor} &= C_L.\mathit{startOld} - o_s \\
            M.\mathit{endAncestor} &=  C_L.\mathit{endOld} \\
            M.\mathit{startLeft} &= C_L.\mathit{startNew} - o_s \\
            M.\mathit{endLeft} &= C_L.\mathit{endNew} \\
            M.\mathit{startRight} &= C_R.\mathit{startNew} \\
            M.\mathit{endRight} &= C_R.\mathit{endNew} + o_e
          \end{align*}
          After the merge region has been recorded, the change $C$ which has a lower $C.\mathit{endOld}$ (or both, if $C_L.\mathit{endOld} = C_R.\mathit{endOld}$) is removed from the list of changes.
    \item If none of the previous cases apply (both $C_R$ and $C_L$ are the same change), both changes are removed from the list of changes (it is a false conflict).
    \item  \label{item:merge_case_empty_right} If the right file does not have any changes left, the remaining changes $C_L$ can be applied without a conflict.
          To compute the position of the merge region $M$ in the right file, the file lengths have to be taken into account (this is analogous to the next change in case \ref{item:merge_case_1}).
          Thus, $M$ is computed as follows:
          \begin{align*}
            M.\mathit{startAncestor} &= C_L.\mathit{startOld} \\
            M.\mathit{endAncestor} &= C_L.\mathit{endOld} \\
            M.\mathit{startLeft} &= C_L.\mathit{startNew} \\
            M.\mathit{endLeft} &= C_L.\mathit{endNew} \\
            M.\mathit{startRight} &= C_L.\mathit{startNew} - R.\mathit{lines} - L.\mathit{lines} \\
            M.\mathit{endRight} &= C_L.\mathit{endNew} - R.\mathit{lines} - L.\mathit{lines}
          \end{align*}
          \item If the left file does not have any changes left, the algorithm behaves symmetrically to case \ref{item:merge_case_empty_right}.
\end{enumerate}

% \documentclass{article}

% \usepackage{ifthen}
% \usepackage{tikz}
% \usepackage{caption}
% \usepackage{subcaption}
% \usetikzlibrary{calc,math}
%\begin{document}
\newcommand{\file}[4][true]{%
    \draw (#2,0.5) -- (#2,-#3);
    \ifthenelse{\equal{#1}{true}}{\draw[dashed] (#2,-#3) -- (#2,-#3-1);}{
        \draw (#2-0.07,-#3) -- (#2+0.07,-#3);
    }
    \node at (#2,0.5) [anchor=south] {#4};
}
\newcommand{\graphicscalemerge}[0]{0.95}

\begin{figure}
    \begin{subfigure}[t]{0.3\textwidth}
        \centering
        \begin{tikzpicture}[scale=\graphicscalemerge]

            \file{-2}{3}{$L$}
            \file{0}{3}{$O$}
            \file{2}{3}{$R$}

            \coordinate (ls) at (-2,0);
            \coordinate (le) at (-2,-1);
            \draw (-1.9, 0) -- (-0.1, 0) -- (-0.1, -1) -- (-1.9, -1) -- cycle;
            \node at (-1, -0.5) {$C_L$};

            \draw (0.1, -1.5) -- (1.9, -2)  -- (1.9, -2.5) -- (0.1, -2.5) -- cycle;
            \node at (1, -2.125) {$C_R$};

            \draw[dashed] (0, 0) -- (2, -0.5);
            \draw[dashed] (0, -1) -- (2, -1.5);

            \draw[ultra thick] (-2, 0) -- (-2, -1);
            \draw[ultra thick] (0, 0) -- (0, -1);
            \draw[ultra thick] (2, -0.5) -- (2, -1.5);
        \end{tikzpicture}
        \caption{Unconflicting changes}
        \label{fig:merge_case_simple}
    \end{subfigure}
    \hfill
    \begin{subfigure}[t]{0.3\textwidth}
        \centering
        \begin{tikzpicture}[scale=\graphicscalemerge]
            \file{-2}{3}{$L$}
            \file{0}{3}{$O$}
            \file{2}{3}{$R$}

            \draw (-1.9, -0.5) -- (-0.1, -0.5) -- (-0.1, -1.5) -- (-1.9, -1.5) -- cycle;
            \node at (-1, -1) {$C_L$};
            
            \draw (0.1, 0) -- (1.9, -0) -- (1.9, -0.5) -- (0.1, -1) -- cycle;
            \node at (1, -0.35) {$C_R$};

            \draw[dashed] (-2, 0) -- (0, -0);
            \draw[dashed] (0, -1.5) -- (2, -1);

            \draw[ultra thick] (-2, 0) -- (-2, -1.5);
            \draw[ultra thick] (0, 0) -- (0, -1.5);
            \draw[ultra thick] (2, -0) -- (2, -1);
        \end{tikzpicture}
        \caption{Conflict}
        \label{fig:merge_case_conflict}
    \end{subfigure}
    \hfill
    \begin{subfigure}[t]{0.3\textwidth}
        \centering
        \begin{tikzpicture} [scale=\graphicscalemerge]
            \file{-2}{3}{$L$}
            \file[ended]{0}{1.9}{$O$}
            \file[ended]{2}{2.4}{$R$}

            \draw (-1.9, 0) -- (-0.1, 0) -- (-0.1, -1) -- (-1.9, -1) -- cycle;
            \node at (-1, -0.5) {$C_L$};
            
            \draw[dashed] (0, -0) -- (2, -0.5);
            \draw[dashed] (0, -1) -- (2, -1.5);

            \draw[ultra thick] (-2, 0) -- (-2, -1);
            \draw[ultra thick] (0, 0) -- (0, -1);
            \draw[ultra thick] (2, -0.5) -- (2, -1.5);

            \node[anchor=north] (eof1) at (0, -2) {EOF};
            \node[anchor=north] (eof2) at (2, -2.5) {EOF}; 

        \end{tikzpicture}
        \caption{Unconflicting changes at the end of the file}
        \label{fig:merge_case_end}
    \end{subfigure}

    \caption{The different cases of the three-way merge algorithm. 
    The figure uses the same way of representing a change hunk as in \Cref{fig:hunk_example}.
    The dashed lines visualize how the position in the other file is computed. 
    The ranges of the final merge hunk $M$ are shown with bold lines.
    It can be seen how the offset of the hunk on the other side (or end of the file) has to be taken account when computing where in the other file a change has to be applied.} 
\label{fig:merge}

\end{figure}
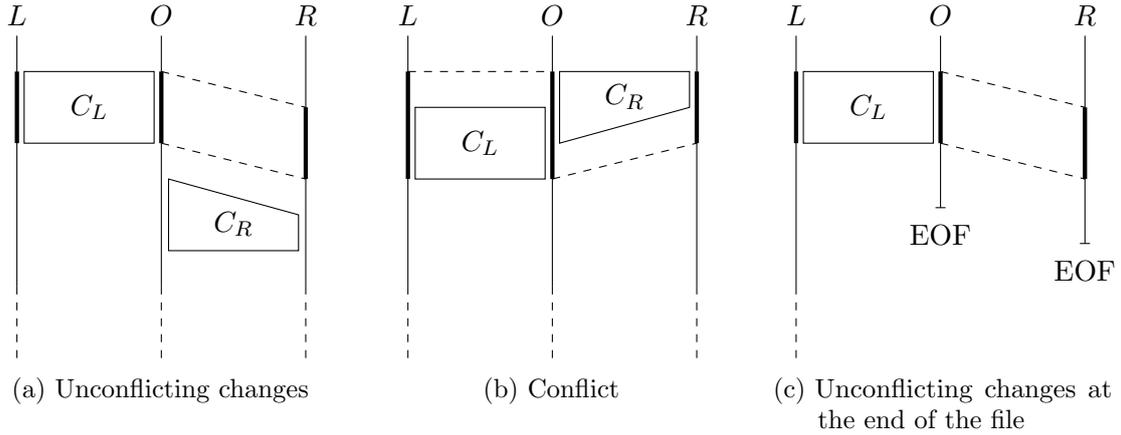

%\end{document}
After no more changes are left, the algorithm computes the merged file by iterating over the merge regions.
Notice that because the edit scripts were generated from diffs, which do not track rearrangements, the merge regions are ordered in all three files.
This makes it simple to generate the merged file by iterating over the merge regions.
If a merge region is a conflict (it originates from case \ref{item:merge_case_conflict}), the conflict is formatted using conflict markers; otherwise, the changed side is copied.

\subsection{Git's Implementation}
While \Cref{section:merge_core_algorithm} described the core three-way merge algorithm as implemented in Git, there are a few extensions that are used on top of it.

\paragraph{Conflict Styles}
When outputting a conflict, by default the content of the merge region $M$ is printed from the left and right files only, separated by the delimiters \texttt{<<<<<<<}, \texttt{=======}, and \texttt{>>>>>>>}.
With the \texttt{merge.conflictStyle=diff3} option, which can also be set from the command line, the region of the ancestor is printed as well, with the additional delimiter \texttt{|||||||}.
This can help with manually resolving the conflicts by giving more context about the changes \cite[p. 279]{chaconProGit2014}.
Using this option does not change the core algorithm, but only the output format.
Furthermore, Git has options to instead of outputting a conflict, always choose either the left, right, or both regions after each other.

\paragraph{Zealous Merges}
By default, Git tries to simplify the conflicts before outputting them.
To do this, it creates a full two-way diff between the merge region in the left and right files \sourcecite{xmerge.c}{363}.
The conflict is then split up such that each hunk in this diff is shown as an individual conflict.
This post-processing simplifies the conflict if a similar but not completely equal change has been done in both files.
In edge cases (see \Cref{section:unnecessary_conflicts} for details), this step can also realize that both sides are equal, removing the conflict completely \commitcite{22b6abc}.
After the conflicts have been split, Git merges all conflicts which have fewer than three non-conflicting lines between them to avoid too many individual conflicts \sourcecite{xmerge.c}{472}.
This analysis is not compatible with the \textit{diff3} output style.
However, there is a \texttt{zdiff3} output style to refine conflicts in that case \sourcecite{xmerge.c}{656}.
Instead of running a full diff between the two regions, it just removes common lines at the start and end of all three regions of the conflict.
This never splits a conflict into multiple conflicts.

\subsection{Related Work}
Similar to finding differences between two files, a three-way merge is not well-defined.
While for diffs there exists the metric of the length of the longest common subsequence (or inversely, the number of lines changed), the merge is a greedy algorithm that does not optimize for such a goal.
For diffs, one can define \textit{correctness} as the property that applying the diff results in the file used to compute the diff.
Such a property is not as clear for merges.
It is already highly context-dependent whether a conflict should be output in a particular case.
Because the algorithm itself is widely used, but little understood, it makes sense to take a look at specific cases and how the algorithm resolves them.
In particular, finding cases where the algorithm behaves in unexpected ways can help both to understand the algorithm better and to gain a better understanding of what a merge should do.

\citeauthor{khannaFormalInvestigationDiff32007} \cite{khannaFormalInvestigationDiff32007} have given a description of the algorithm and analyzed some cases formally.
Their description is slightly more abstract than the one given in \Cref{section:merge_core_algorithm} and only contains the core algorithm without the extensions like zealous merging.
In particular, their description of the algorithm represents the diffs as an arbitrary \textit{maximum matching} between unchanged lines.
While every \textit{minimal} diff corresponds to a maximum matching, not every maximum matching is even a valid diff (diffs do not allow rearrangements).
Furthermore, the diff algorithm used by default for merges in Git does not generate minimal diffs, by design (as seen in \Cref{section:diffs}).
They also make no assumption about \textit{which} maximum matching is used, ignoring the fact that the diff algorithms produce diffs with properties like the LCS of unique lines in the case of the patience algorithm.
Since the diffs have a large impact on the results of the merge (which we will see below), this is a significant difference.

They observe that a common intuition about merges is that, if two changes are separated by a large enough region of unchanged lines, they will not conflict, which they formalize as \textit{locality}.
This property does not hold in general, however.
Suppose that a file contains the lines \texttt{abab}. 
On one side, the two lines \texttt{ab} are added to the front, resulting in the file \texttt{ababab}.
On the other side, the last two lines are removed and replaced with a \texttt{c}, resulting in the file \texttt{ababc}.
Since on one side the beginning of the file was changed and on the other side the end, one would not expect a conflict.
However, Git creates the following two diffs for running a three-way merge: \texttt{abab\textcolor{diffincl}{\{+ab+\}}} and \texttt{abab\textcolor{diffrem}{\{-ab-\}}\textcolor{diffincl}{\{+c+\}}}.
These two diffs do result in a conflict because both diffs modify the end of the file.
This example works with an arbitrary number of \texttt{ab} lines in the file, showing that even if the two changes are separated by a large number of unchanged lines, the merge can still conflict.

They are able to show that it can be guaranteed that no conflict occurs, if a line that is unique in all three versions is between the two changes, however.

\subsection{Effect of the Diff Algorithms}
\label{section:merge_evaluation}

Git's \texttt{ort} merge strategy, which is the default since version 2.34, started to default to using the histogram diff algorithm instead of the Myers algorithm.
To evaluate the effect of this change, this section analyzes empirically how the merge algorithm behaves with the different diff algorithms.
I have run an empirical evaluation because, as described above, few theoretical properties of a ``good'' merge exist.
The later sections will focus on individual examples, analyzing how the merge behaves in detail.

\paragraph{Methods}
To get a source of representative merges from real software development, the merge commits in a Git repository can be used.
For this analysis, I extracted individual three-way merges from the merge commits in the history of Git itself.
However, merge commits do not document the entire process by which the merge was done; they only record the parent commits and the result.
When trying to extract individual three-way merges from merge commits, multiple issues arise, among others:
\begin{enumerate}
    \item Merge commits do not store which files exactly were merged (due to e.g., rename detection).
    \item Merge commits do not record which strategy was used and which merge base was selected.
    \item The merges might have been done with a different version of Git.
    \item The result might have been manually modified after the merge.
\end{enumerate}
To get meaningful results about the impact of the diff algorithms, it is not necessary to replay the exact merges as they were done.
While not perfect, approximating these merges still provides potential merges of real-world source code that can be analyzed.
Concretely, I have only considered merge commits with exactly two parents, recovered the merge base using the \texttt{git merge-base} command (ignoring the commit if there are multiple), and matched files in the left, right, and ancestor versions only by their names.
I have not used the content of the merge commits (which might contain the manual conflict resolution) to evaluate the merges since it is not clear whether all conflicts were resolved correctly in the merge commit or fixed with a later commit.
Furthermore, a simple comparison with the content of the merge commit would favor the diff algorithm that was used originally.

\paragraph{Results}
Out of the 19,888 merge commits in the history of Git at version 2.49, 12,625 contained three-way merges in line with the above criteria.
This resulted in 57,050 individual merges of three files.
Out of these, about 28\% had conflicts with one of the algorithms, while the other roughly 72\% of merges did not result in conflicts with any of the algorithms.
The following analysis will primarily focus on the merges which did result in conflicts, since this is where a merge and its quality become interesting.
However, no conflict being detected does not imply that it was clear how the merge should be done.
In four cases, no conflict occurred with any of the four algorithms; however, the result was different between the diff algorithms used.
This effect is explained in more detail in \Cref{section:duplicated_changes}.

Possible metrics to evaluate a merge by could be (1) the number of conflicts or (2) the number of conflicting lines in a file.
Since two smaller conflicts might be more desirable than a single large conflict (zealous merges even split conflicts by design), the former is not as clearly positive as it seems at first.
Therefore, I used the total number of conflicting lines in a file.
This is not a perfect metric, since it does not take the content into account, but can be used to get a high-level overview of the merge quality.
I have also chosen this metric because, in my qualitative research, the effect outlined in \Cref{section:unnecessary_conflicts} (conflicts where the resolution seems clear) was the most common issue with merges.

\begin{figure}
    \centering
    \includegraphics[width=0.9\textwidth]{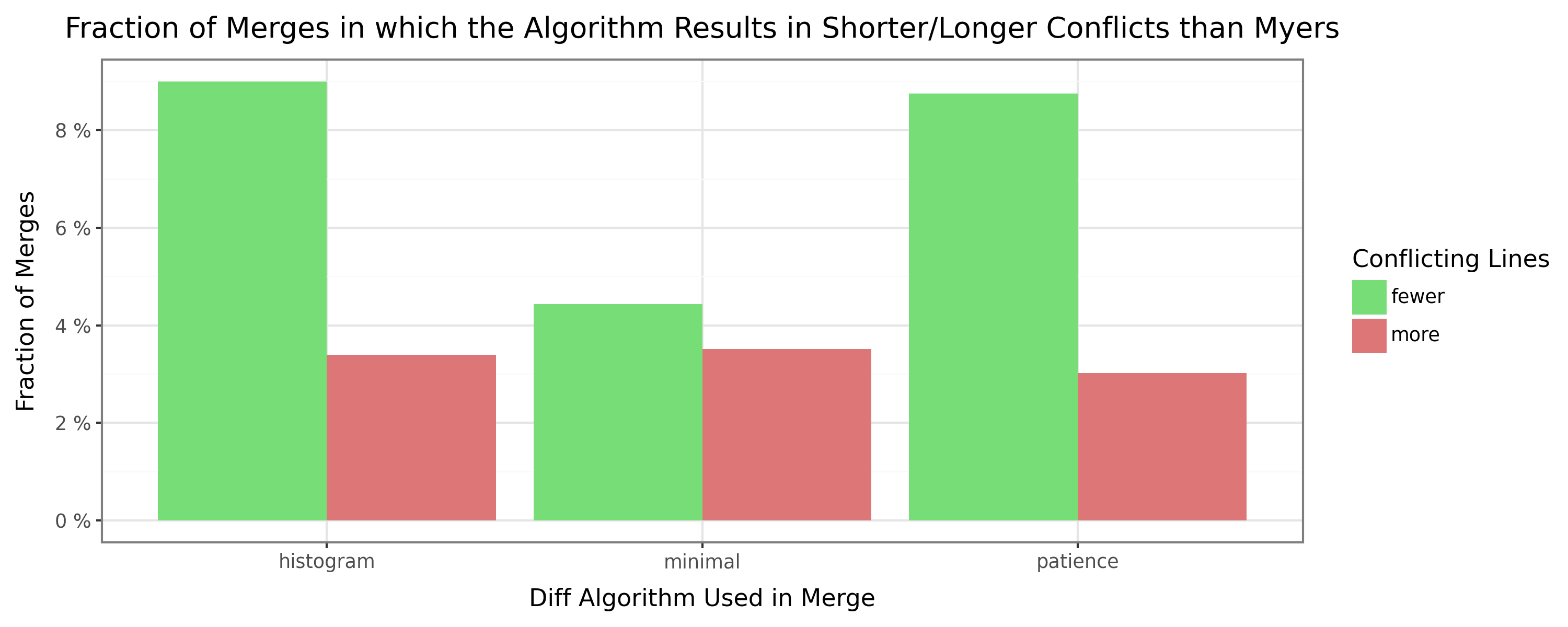}
    \caption{Comparison of conflict sizes when using different diff algorithms.
        The graphics shows, for real merges from the history of Git, whether merging with
        the respective algorithm results in a larger or smaller conflict than using the Myers algorithm.
        Both the histogram and patience algorithms result in smaller conflicts in almost 10\% of merges.}
    \label{fig:merge_diff_algorithms}
\end{figure}

The total number of conflicting lines varies greatly and can be very large in some files (e.g., auto-generated files).
To get a simpler metric of how complex the conflicts are, I compared merges using the histogram, patience, and minimal algorithms to merges using the Myers algorithm for the underlying diffs.
I only count whether the merge results in more or fewer conflicting lines than the merge using the Myers algorithm.
The results of this are shown in \Cref{fig:merge_diff_algorithms}.
All three algorithms (minimal, patience, and histogram) result in longer conflicts than Myers in about 3\% of merges with conflicts.
The minimal algorithm also has shorter conflicts than Myers in about 3\% of merges.
This indicates that \texttt{myers} and \texttt{minimal}, while not always giving the same results, are roughly equally good for use in the merge algorithm.
In contrast, the patience and histogram algorithms result in shorter conflicts in almost 10\% of merges with conflicts.
This result supports the use of the histogram algorithm when merging, as is the default in Git.
An intuition for why this might be the case is given in \Cref{section:unnecessary_conflicts}.

\section{False Conflicts}
\label{section:unnecessary_conflicts}
A common pattern that I observed when analyzing the results of the experiment in \Cref{section:merge_evaluation} was that multiple conflicts are output when both files changed a part of a file in exactly the same way (which should have been a false conflict and just applied).
I have observed this behavior more frequently when using the Myers and minimal algorithms than with the other two.

The reason why these unnecessary conflicts occur is because, while both sides have changed a part of the file in exactly the same way, a different diff is generated for the two sides.
While the diff algorithms are deterministic (when the entire input files are the same, the output will always be the same), a small change at the beginning of a file might cause the algorithm to choose a different diff further down in the file.
Unfortunately, this makes these examples difficult to simplify while still keeping them working with the exact algorithm of Git.
However, consider the following example, which illustrates the effect.
A part of a file $O =$ {\verb|[..]XaY[..]|} is changed on both sides ($L$ and $R$) to \verb|[..]XaaY[..]|.
There are other changes in the omitted parts of the files, making $L \neq R$.
If in $L$ the first \texttt{a} is marked as added, while in $R$ the second \texttt{a} is marked as added, the three-way merge algorithm will generate a conflict since the diff hunks do not match.
The conflict block will contain \texttt{aa} on both sides, which should not have been marked as a conflict.
These simple cases are almost always automatically detected by Git and filtered out when running the simplification for zealous merges (this was previously a bug \commitcite{22b6abc}).
However, in more complex examples (in particular with two such changes close to each other), the sides do not always match and conflicts are emitted.

\paragraph{Remaining False Conflicts}
When running the experiment, I checked whether Git always detects false conflicts and never emits conflicts where both sides are the same.
As mentioned, this behavior was previously a bug and was supposed to be fixed.
Interestingly, I have found an edge case where Git still outputs conflicts which are equal on both sides.
During the evaluation, this was observable in about 1000 merges (2\% of merges with conflicts) with the Myers algorithm and in about 500 merges (0.7\% of merges with conflicts) with the histogram algorithm.

Only after the core procedure of zealous merging is done (diffing $L$ and $R$ and splitting the conflict or detecting a false conflict), conflict hunks that have fewer than three lines in between are merged together.
In this step, it can also happen that both sides of the conflict become equal, which is not noticed by Git.
This can be considered a bug, and I am planning on developing a fix and submitting it to the Git project.
The issue can be solved by re-checking for false conflicts after the conflict merging.
While this is not a fundamental issue with the core algorithm and more an implementation issue, it does demonstrate that implementing the three-way merge has great potential to create subtle problems.

\section{Duplicated Changes and Missing Conflicts}
\label{section:duplicated_changes}
\begin{figure*}
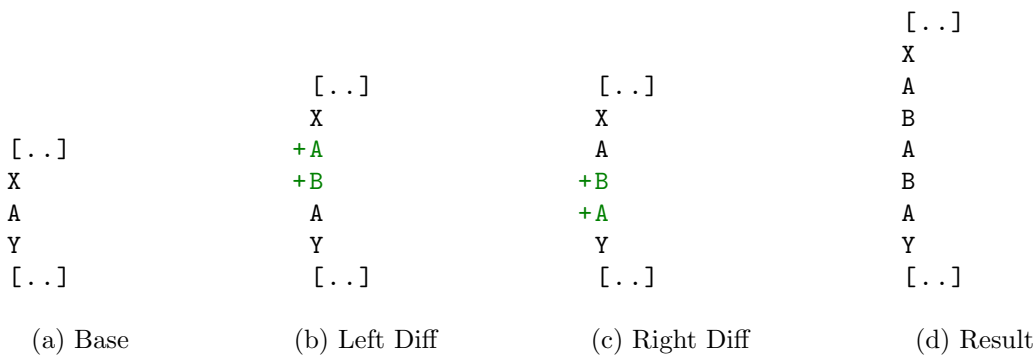

    \centering
    \subcaptionbox{Base}[2cm]{%
\begin{lstlisting}[language=diff, basicstyle=\small\ttfamily]
[..]
X
A
Y
[..]
\end{lstlisting}}
    \hspace{1.5cm}
    \subcaptionbox{Left Diff}[2cm]{%
\begin{lstlisting}[language=diff, basicstyle=\small\ttfamily]
 [..]
 X
+A
+B
 A
 Y
 [..]
\end{lstlisting}}
    \hspace{1.5cm}
    \subcaptionbox{Right Diff}[2.5cm]{%
\begin{lstlisting}[language=diff, basicstyle=\small\ttfamily]
 [..]
 X
 A
+B
+A
 Y
 [..]
\end{lstlisting}}
    \hspace{1.5cm}
    \subcaptionbox{Result}[2cm]{%
\begin{lstlisting}[language=diff, basicstyle=\small\ttfamily]
[..]
X
A
B
A
B
A
Y
[..]
\end{lstlisting}}

    \caption{Duplicated addition due to inconsistent diffs.
        Notice that while the diffs are different, the left and right versions of the file are identical.
        When the three-way merge is run on these, it detects that the \texttt{A} can be used as a common line between the two changes and sees the situation as two distinct additions.}
    \label{fig:duplicated_addition}
\end{figure*}

Similar to how a section of a file being diffed differently despite being the same change resulted in unnecessary conflicts in the previous section, the same can also have an arguably worse effect: duplicating a change without causing a conflict.
I have created an example to illustrate the effect in \Cref{fig:duplicated_addition}.
In that case, the section of ancestor file $O$ \texttt{[..]XAY[..]} is changed on both sides ($L$ and $R$) to \texttt{[..]XABAY[..]}.
Since the same change happened on both branches, one would expect that the result of the three-way merge is simply \texttt{[..]XABAY[..]}.
However, if one of the two diffs selects \texttt{AB} as added and the other selects \texttt{BA} as added (both equally valid, minimal diffs), the three-way merge will not recognize that the two changes are in fact the same.
Unlike in the previous section, this does not result in a conflict, but instead the addition is duplicated without a conflict being emitted, because the other \texttt{A} separates the two changes.
While this is already highly unintuitive, the same effect can not just happen when both sides add the same text, but also when the changes are different.
In the same example, if the left side is changed to \texttt{[..]XABAY[..]} and the right side to \texttt{[..]XACAY[..]}, it looks like the most simple case of a conflict.
One side added a \texttt{B} and the other side added a \texttt{C}.
However, it can happen that, due to inconsistent diffs, the result is \texttt{[..]XABACAY[..]} without a conflict being emitted.

The two examples are not just theoretical. 
In fact, I have noticed them when inspecting the results of the experiment in \Cref{section:merge_evaluation}.
The four cases where no conflict occurred, but the result was different between the algorithms, were all caused by this effect.
Suppose in the example of \Cref{fig:duplicated_addition} \texttt{X}, \texttt{B}, and \texttt{Y} are functions while \texttt{A} is an empty line.
The example is then simply the very common case of adding a function, with empty lines between them.
The problem \textit{always} occurs if one diff marks the empty line before the function and one the empty line after the function as added.

\section{Commutativity of Merges}
One might expect merges to be commutative, namely that merging branch $A$ into branch $B$ results in the same files as merging branch $B$ into branch $A$ (up to the order inside the conflict blocks).
The entire algorithm---as described in \Cref{section:merge_core_algorithm}---is symmetric.
At no point is a distinction made between the two files and all cases exist for both files symmetrically.
The only non-symmetric part is that the text between merge regions is copied from the left file.
However, these parts are always unchanged in both files and therefore equal.

Despite this, I have noticed that merges are not commutative in general by default in Git.
\begin{figure*}
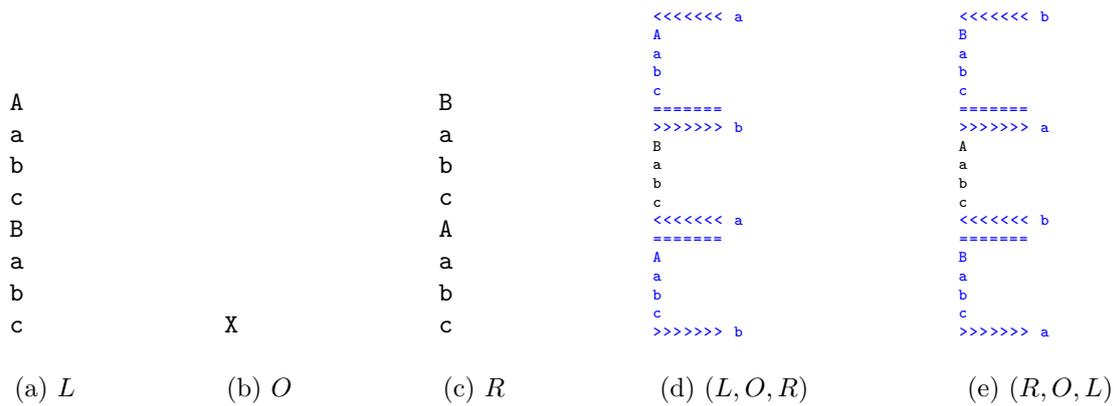

    \subcaptionbox{$L$}[1cm]{%
\begin{lstlisting}[language=diff, basicstyle=\small\ttfamily]
A
a
b
c
B
a
b
c
\end{lstlisting}}
    \hfill
    \subcaptionbox{$O$}[1cm]{%
\begin{lstlisting}[language=diff, basicstyle=\small\ttfamily]
X
\end{lstlisting}}
    \hfill
    \subcaptionbox{$R$}[1cm]{%
\begin{lstlisting}[language=diff, basicstyle=\small\ttfamily]
B
a
b
c
A
a
b
c
\end{lstlisting}}
    \hfill
    \subcaptionbox{$(L,O,R)$\label{fig:merge_commutative_aob}}[0.15\textwidth]{%
\begin{lstlisting}[language=diff, basicstyle=\tiny\ttfamily]
<<<<<<< a
A
a
b
c
=======
>>>>>>> b
B
a
b
c
<<<<<<< a
=======
A
a
b
c
>>>>>>> b
\end{lstlisting}}
    \hfill
    \subcaptionbox{$(R,O,L)$\label{fig:merge_commutative_boa}}[0.15\textwidth]{%
\begin{lstlisting}[language=diff, basicstyle=\tiny\ttfamily]
<<<<<<< b
B
a
b
c
=======
>>>>>>> a
A
a
b
c
<<<<<<< b
=======
B
a
b
c
>>>>>>> a
\end{lstlisting}}
    \caption{Example of a non-commutative merge.
        This example depends on the histogram diff algorithm.
        Note that the merges $(L,O,R)$ and $(R,O,L)$ differ not only in the order of the conflict blocks, but also in the common text between the two conflicts.}
    \label{fig:merge_commutative}
\end{figure*}
This is caused by the conflict refinement step when using zealous merges (which is the default).
The core step of zealous merges runs a two-way diff between the left and right files.
This is done with the same diff algorithm used for the rest of the merge (by default the histogram algorithm).
In \Cref{section:diffs}, I have shown that the histogram algorithm is not commutative (\Cref{fig:histogram_bad} is an example of this).
This means that the two-way diff depends on the order of the files. 
Swapping $L$ and $R$ might make the histogram algorithm select a different pivot and produce a different diff.

I have created the example in \Cref{fig:merge_commutative} to demonstrate that this is not just a theoretical issue, but can actually result in a non-commutative merge.
The difficulty in creating such an example is that the zealous merge only starts to take effect when at least three unchanged lines are part of the diff.
In the example, the ancestor file only contains a single line, \texttt{X}.
In one branch, the file is changed to $L = $\texttt{AabcBabc} and in the other branch to $R = $\texttt{BabcAabc} (the \texttt{A} and \texttt{B} are swapped and the \texttt{abc} are used to reach three unchanged lines).
One might expect that the diff between $L$ and $R$ simply adds and removes the uppercase letters to swap them, keeping the lowercase letters unchanged.
However, due to the lowercase letters being more frequent, the histogram algorithm chooses to match the first uppercase letter of the left file with its counterpart in the right file, marking the lowercase letters in between as changed.
This depends on the order of the files: in this setup, the algorithm always selects the first unique letter of the left file as the pivot.
As a result, the merge $(L, O, R)$ shows a \texttt{B} between the two conflicts, while the merge $(R, O, L)$ shows an \texttt{A} between the two conflicts.

\section{Commutativity of Rebases}
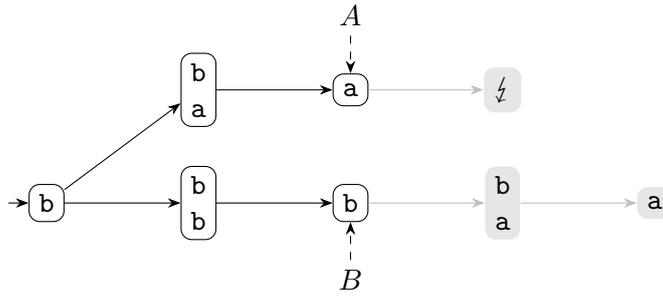
\begin{figure*}
    \centering
    \begin{tikzpicture}[
            file/.style={rectangle, draw=black, rounded corners, font=\ttfamily, align=left},
            rebased/.style={rectangle, draw=none, rounded corners, font=\ttfamily, align=left, fill=gray!20},
            >=Stealth
        ]
        \node[file] (o) at (0,0) {b};
        \node[file] (a1) at (2, 0) {b\\b};
        \node[file] (a2) at (4, 0) {b};
        \node[file] (b1) at (2, 1.5) {b\\a};
        \node[file] (b2) at (4, 1.5) {a};
        \node[above =5mm of b2] (A) {$A$};
        \node[below=5mm of a2] (B) {$B$};
        \draw[->, dashed] (A) -- (b2);
        \draw[->, dashed] (B) -- (a2);

        \draw[->] (o) -- (a1);
        \draw[->] (a1) -- (a2);
        \draw[->] (o) -- (b1);
        \draw[->] (b1) -- (b2);
        \draw[->] (-0.5, 0) -- (o);

        \node[rebased] (a3) at (6, 0) {b\\a};
        \node[rebased] (a4) at (8, 0) {a};
        \draw[->, draw=gray!50] (a2) -- (a3);
        \draw[->, draw=gray!50] (a3) -- (a4);

        \node[rebased] (b3) at (6, 1.5) {$\lightning$};
        \draw[->, draw=gray!50] (b2) -- (b3);
    \end{tikzpicture}
    \caption{Minimal non-commutative rebase. Rebasing $B$ on $A$ results in a conflict while rebasing $A$ on $B$ can finish without a conflict.}
    \label{fig:commutative_rebase}
\end{figure*}

Similar to the commutativity of merges, one might expect rebases to be commutative, meaning that rebasing branch $A$ on branch $B$ results in the same files as rebasing branch $B$ on branch $A$.
This is not the case, as an inherent limitation of running a rebase by repeated cherry-picking through three-way merges.
I have found a counterexample by ``fuzzing'' Git, repeatedly checking whether rebases are symmetric with random files.
\Cref{fig:commutative_rebase} shows a minimal example where it is not the case.
The file contents are shown directly in the commits.
When rebasing the branch $A$ on top of $B$, both cherry-picks can finish without a conflict.
Because the base of $A$ is exactly the same as the last commit of $B$, this rebase is trivial.
When rebasing $B$ onto $A$, however, the first cherry-pick results in a merge conflict.
The cherry-pick will run a three-way merge of \texttt{a} and \texttt{bb}, with \texttt{b} as the base.
This clearly has to result in a conflict (one side removes the \texttt{b}, while the other side adds a second one at the same position).

Generalizing the minimal example, this behavior always occurs when two branches modify the same area, but one branch reverts the changes and the other does not.
In contrast to a merge (which would merge in both directions to \texttt{a}), rebases are affected by the entire history of the branches, not just the two latest commits and the common ancestor.

\chapter{Conclusion}
I have analyzed the text-based collaboration algorithms used in Git, the most widely used version control system, with a focus on its diff and merge procedures and their practical behavior.
Git uses a three-way merge for many of its operations where one might not expect it, like cherry-picking, reverting, rebasing, and even applying patches.
Thus, the three-way merge algorithm (and the diff algorithm it relies on) can be seen as the central algorithm of Git.

Git uses multiple techniques that improve the day-to-day usage in practice.
This includes, but is not limited to, the more intuitive diff algorithms based on line frequency, the post-processing steps that make diffs more readable based on real-world data, and the zealous merging algorithm that simplifies conflicts.
These techniques, while adding some complexity, could be used as inspiration for improvements in other version control systems, or even text-based collaboration in general.

While my experiments have shown that the algorithms perform well in most cases (as would be expected from a system as widely used as Git), there are many edge cases that lead to surprising results.
For both diffs and merges, few formal guarantees can be given, and the algorithms focus on practical behavior rather than theoretical guarantees.
Nevertheless, observations like the exponential time complexity of merges, and the fact that merges and rebases are not commutative, are interesting properties of the system.
For merges in particular, the algorithm provides very few guarantees on the outcome, and many potentially highly unintuitive results can occur.
In the future, it would be interesting to try to create a more formal framework for a merge algorithm, which explicitly specifies which merges should succeed and which should result in a conflict.
Having such a framework could also help improve the algorithm by providing a clearer understanding of its behavior and potential edge cases.
The fact that it is still possible to find bugs that affect the usage of Git shows that a more formal understanding of the merge algorithm could be beneficial.

Overall, this work provides insights into Git's practical design decisions—including frequency-based diff algorithms, three-way merges, and zealous merging—that enable effective collaborative software development.
The unexpected behaviors found further highlight the need for rigorous theoretical foundations to improve reliability in the merge process.

\newpage
\noindent
\printbibliography
\end{document}